\documentclass[twocolumn,showpacs,preprintnumbers,amsmath,amssymb,%
prd,a4paper,floatfix,nofootinbib,superscriptaddress]{revtex4-1}  

\usepackage{amssymb,amsfonts,amsmath,bm}
\usepackage{graphicx}
\usepackage{hyperref}

\makeatletter

\renewcommand{\@makefntext}[1]{\parindent=1em\noindent\hbox to 1.8em{\hss$^{\@thefnmark}$}#1}
\renewcommand{\@footnotemark}{\hbox{\mathsurround=0pt$^{\@thefnmark}$}}

\makeatother

\usepackage[tight]{units}

\newcommand{\Fig}[1]{Fig.~\ref{#1}}
\newcommand{\Tab}[1]{Table~\ref{#1}}
\newcommand{\Sec}[1]{Sec.~\ref{#1}}
\newcommand{\Eq}[1]{Eq.~(\ref{#1})}

\newcommand{\RD}[1]{{\mathrm{red}(#1)}}
\newcommand{\LM}[1]{{\mathrm{lm}(#1)}}
\newcommand{\SU}[1]{\mathrm{SU}(#1)}

\newcommand{\ket}[1]{\left|{#1}\right>}
\newcommand{\bra}[1]{\left<{#1}\right|}
\newcommand{\gaf}{\gamma_5}

\newcommand{\be}{\begin{equation}}
\newcommand{\ee}{\end{equation}}

\newcommand{\kslash}{k\hspace{-2mm}\slash}
\newcommand{\pslash}{p\hspace{-0.475em}\slash}

\newcommand{\tr}[1]{\,\text{tr}\left[#1\right]}
\renewcommand{\det}[1]{\,\text{det}\left[#1\right]}

\renewcommand{\bar}{\overline}

\renewcommand{\Re}{\mathfrak{Re}\,}

\newcommand{\bigO}{\mathcal{O}}
\newcommand{\DCI}{D_{\textrm{CI }}}
\newcommand{\1}{\mathrm{1\hspace{-0.4em}1}}
\def\symbR{~\hspace{-0.5em}\textsuperscript{\tiny{\textregistered}} }
\def\symbTM{~\hspace{-0.5em}\textsuperscript{\tiny{\texttrademark}} }

\begin{document}
\date{\today}

\title{The chirally improved quark propagator and restoration of chiral symmetry}
\author{Mario Schr\"ock}
\email{mario.schroeck@uni-graz.at}
\affiliation{Institut f\"ur Physik, FB Theoretische Physik, Universit\"at
Graz, A--8010 Graz, Austria}

\begin{abstract}
The chirally improved (CI) quark propagator in Landau gauge 
is calculated in two flavor lattice Quantum Chromodynamics.
Its wave-function renormalization function $Z(p^2)$
and mass function  $M(p^2)$ are studied. 
To minimize lattice artifacts,
tree-level improvement of the propagator and tree-level correction 
of the lattice dressing functions is applied.
Subsequently the CI quark propagator under 
Dirac operator low-mode removal is investigated.
The dynamically generated mass in the infrared domain of the mass function
is found to
dissolve continuously as a function of the reduction level
and strong suppression of $Z(p^2)$ for small momenta is observed.
\end{abstract}
\maketitle

\section{Introduction}
The quark propagator is one of the fundamental objects in 
Quantum Chromodynamics (QCD).
The mass function of the quark propagator reveals the value of
the running quark mass in the deep ultraviolet (UV)
where interactions are weak due to the asymptotic freedom of QCD.
In the infrared (IR), the dynamical generation of mass which is 
associated with the spontaneous breaking of chiral symmetry is exhibited
by the mass function.
The IR is not accessible with perturbative methods; lattice QCD
provides a nonperturbative \emph{ab initio} approach to QCD
and thus is a well adapted tool to study the IR physics of the 
strong nuclear force. 

The quark propagator is a gauge dependent object and thus the gauge
has to be fixed in order to study its properties;
we adopt the manifestly Lorentz covariant Landau gauge for the present work.
The Landau gauge quark propagator has been studied on the lattice
with various fermionic actions.
Some initial investigations using (improved) Wilson fermions have been reported in Ref. \cite{Skullerud:2000un, Skullerud:2001aw}.
A series of studies using standard Kogut--Susskind \cite{Kogut:1974ag} and 
Asqtad \cite{Orginos:1999cr}  quarks
found that staggered quarks are well suited to explore the properties of
the quark propagator on the lattice \cite{Bowman:2002bm, Parappilly:2005ei, Bowman:2001xh, Bowman:2002kn, Bowman:2005vx, Furui:2005mp}.

Lattice Dirac operators that fulfill the Ginsparg--Wilson (GW) equation
allow for lattice fermions that have an exact chiral symmetry
at nonzero lattice spacing.
The overlap operator \cite{Neuberger:1997fp,Neuberger:1998wv} 
provides a solution to the GW equation. The quark propagator from the
overlap action has been examined in
 \cite{Bonnet:2002ih, Zhang:2003faa, Zhang:2004gv, Kamleh:2004aw, Kamleh:2007ud, Bowman:2005zi, Bowman:2004xi}.
The drawback of overlap fermions is their very high computational cost
which renders them impractical for full dynamical simulations.

In this letter we analyze the quark propagator from the so-called chirally 
improved (CI) Dirac operator \cite{Gattringer:2000js, Gattringer:2000qu} which fulfills the GW equation
not exactly, but only approximately.
Nevertheless, the gain in simulation time of roughly one order of magnitude, 
in comparison to overlap fermions, allows for an investigation of the propagator 
on full dynamical configurations \cite{Gattringer:2008vj, Engel:2010my}.
The better chiral properties of the CI operator as opposed to Wilson's fermion action
make it well suited to 
explore effects of spontaneous chiral symmetry breaking on the lattice.

Banks and Casher formulated a relation of the density
of the smallest nonzero eigenvalues of the Dirac operator to the chiral 
condensate \cite{Banks:1979yr}.
In \cite{Lang:2011vw} we have studied the effects of removing the lowest eigenmodes
of the Hermitian CI Dirac operator $\gaf\DCI$ on the meson spectrum 
and found signals for the restoration of chiral symmetry 
(the masses of the
$\rho$ and $a_1$ became approximately degenerate, cf. \cite{Glozman:2007ek}) whereas confining properties persisted.
The authors of \cite{Suganuma:2011kn} expand the Wilson loop in terms of Dirac operator
eigenmodes and detect that removing the lowest modes does not influence the static
quark potential qualitatively.

A portion of this study aims at answering the question, 
how change the quark wave-function renormalization function
$Z(p^2)$ and the quark mass function $M(p^2)$ under Dirac low-mode removal?
It is expected that the mass function flattens out in the IR once chiral
symmetry is restored.
Yet another question of interest is how
the Dirac eigenmode truncation level at which
chiral symmetry was found to be approximately restored \cite{Lang:2011vw},
compares to the loss of dynamical mass generation in $M(p^2)$
as a function of the truncation level.

The remainder of this work is as follows: in \Sec{sec:gaugefix} we briefly
summarize the defining equations of lattice Landau gauge fixing.
In \Sec{sec:CIquark} we first remind the reader of the main steps in the construction 
of the $\DCI$ operator,  followed by a discussion of $Z(p^2)$ and $M(p^2)$ from the 
$\DCI$ at tree-level and in the full interacting case.
In order to reduce the dominant lattice artifacts we apply tree-level improvement
and test a multiplicative and an hybrid scheme of tree-level correction.
In \Sec{sec:restoration} we investigate $Z(p^2)$ and $M(p^2)$ from the 
$\DCI$ under Dirac low-mode removal and in 
\Sec{sec:conclusions} we summarize and conclude.

\section{Gauge fixing}\label{sec:gaugefix}

The continuum Landau gauge condition,
\be
	\partial_\mu A_\mu(x) = 0,
\ee
can be realized on the lattice by requiring the maximization of
the gauge functional
\be\label{gaugefunc}
	F_g[U] = \Re\sum_{ \mu, x} \mathrm{tr}\big[ U^g_ \mu(x) + U^g_ \mu(x-\hat \mu)^\dagger\big]
\ee
with respect to gauge transformations $g(x)\in\mathrm{SU}(3)$ where
\be
 U^g_\mu(x) \equiv g(x) U_\mu(x) g(x+\hat\mu)^\dagger.
\ee
The sum in \Eq{gaugefunc} runs over the four Dirac components $\mu$ and all lattice sites $x$.
Once such a gauge transformation is found,
the discrete Landau gauge condition
\be
	\Delta(x)\equiv \sum_\mu\left(A_\mu(x)-A_\mu(x-\hat \mu) \right) =0
\ee
holds, where $A_\mu(x)$ is recovered from the lattice gauge links $U_\mu(x)$ via
\be
	A_\mu(x)\equiv \left[\frac{U_\mu(x)-U_\mu(x)^\dagger}{2iag_0}\right]_{\textrm{traceless}}.
\ee
A measure for the achieved Landau gauge ``quality''
is given by 
\be\label{gaugeprec}
	\theta \equiv \frac{1}{VN_c}\sum_{x}\tr{\Delta(x)\Delta(x)^\dagger},
\ee
here the trace goes over the color indices, $N_c$ is the number of colors 
and $V$ is the number of lattice points. 
In the later discussion of the CI quark propagator we will
choose $\theta< 10^{-10}$ as the stopping criterion for the gauge fixing algorithm.

We accelerate the costly task of lattice gauge fixing by utilization
of the graphics processing unit (GPU) with 
NVIDIA\symbR's CUDA\symbTM (Compute Unified Device Architecture) programming environment,
as pointed out in the Appendix \ref{sec:GPU}.

For a general discussion of lattice gauge fixing and its problems we refer to \cite{Giusti:2001xf}.

\section{The CI quark propagator}\label{sec:CIquark}
In the present section we analyze the lattice dressing functions from the
CI quark propagator after having repeated the main steps in the construction 
of the CI Dirac operator.

\subsection{The CI Dirac operator}
The so-called chirally improved Dirac operator $\DCI$ was introduced 
in \cite{Gattringer:2000js} and first analyzed in \cite{Gattringer:2000qu}
where also its spectral properties were studied.
An initial quenched hadron spectroscopy using the $\DCI$ was examined in \cite{Gattringer:2003qx}
before dynamical configurations including two light degenerate CI quarks have been generated
in order to calculate the hadron spectrum in a series of papers \cite{Lang:2005jz, Gattringer:2008vj, Engel:2010my, Engel:2011zr}.
Renormalization factors of quark bilinears of the $\DCI$ were studied in \cite{Gattringer:2004iv, Huber:2010zza}.

The CI Dirac operator is an approximate solution to the GW equation.
It is constructed by expanding the most general Dirac operator in a basis of simple
operators, 
\be\label{DCIansatz}
	\DCI(x,y) = \sum_{i=1}^{16} c_{xy}^{(i)}(U)\Gamma_i + m_0\1,
\ee
where the sum runs over all elements $\Gamma_i$ of the Clifford algebra.
The coefficients $c_{xy}^{(i)}(U)$ consist of path ordered products of the link variables $U$
connecting lattice sites $x$ and $y$.
Inserting this expansion into the GW equation
then turns into a system of coupled quadratic equations for the
expansion coefficients of the $\DCI$.
That expansion provides for a natural cutoff which turns the quadratic equations
into a simple finite system.

The ansatz is constructed such that all symmetries of the 
fermionic action are maintained and moreover $\gamma_5$-hermiticity is imposed.
The so-called clover term \cite{ShWo85} is included
for $\bigO(a)$ improvement where the $c_{\textrm{sw}}$ parameter is set to its tree-level value (one).

\subsection{Configurations}
For the analysis of the CI quark propagator we use 125 gauge field configurations
\cite{Gattringer:2008vj, Engel:2010my} of  lattice size $16^3\times 32$ and lattice spacing $a=\unit[0.144(1)]{fm}$.
The configurations include two light degenerate dynamical CI quark flavors with the
mass parameter set to $m_0=-0.077$ and a resulting bare AWI-mass of $m=\unit[15.3(3)]{MeV}$.
For the simulation of the gauge fields as well as for our valence quarks, paths up to length four 
are used in the ansatz \Eq{DCIansatz} and the corresponding coefficients can be found in \cite{Gattringer:2008vj}.

\subsection{Nonperturbative quark propagator}
The continuum quark propagator at tree-level reads
\be\label{treelevelcont}
	S^{(0)}(p) = \left(i\pslash +m\right)^{-1}
\ee
where $m$ is the bare quark mass.
In a manifestly covariant gauge like Landau gauge, the interacting 
renormalized quark propagator $S(\mu;p)$
can be decomposed into Dirac scalar and vector parts
\be
	S(\mu;p) = \left(i\pslash A(\mu;p^2)+B(\mu;p^2)\right)^{-1}
\ee
or equivalently as
\be\label{fullcont}
	S(\mu;p) = Z(\mu;p^2)\left(i\pslash +M(p^2)\right)^{-1}.
\ee
In the last equation we introduced
the wave-function renormalization function $Z(\mu;p^2)=1/A(\mu;p^2)$ and
the mass function $M(p^2)=B(\mu;p^2)/A(\mu;p^2)$.

On the lattice, the regularized quark propagator is calculated
and consequently it depends on the cutoff $a$. The regularized quark propagator
$S_L(p;a)$ can then be renormalized at the renormalization point $\mu$ 
with the momentum independent quark wave-function renormalization constant $Z_2(\mu;a)$,
\be
	S_L(p;a) = Z_2(\mu;a) S(\mu;p).
\ee

Whereas the mass function $M(p^2)$ is independent of the renormalization point $\mu$ 
(and equivalently of the cutoff scale $a$), the wave-function renormalization function $Z(\mu;p^2)$
differs at different scales but can be related from different scales
by multiplication with a constant, i.e., by the ratio of the two different
quark renormalization constants.

The momentum subtraction scheme (MOM) has the renormalization point boundary conditions
$Z(\mu;\mu^2)=1$ and $M(\mu^2)=m(\mu)$ where $m(\mu)$ becomes the running mass at large momenta.

Below we extract the nonperturbative functions $M(p^2)$ and 
$ Z(p^2) \equiv Z_2(\mu;a)Z(\mu;p^2)$ directly
from a lattice calculation as it was discussed in great detail in, e.g., 
Ref. \cite{Skullerud:1999gv}.
We perform a cylinder-cut \cite{Skullerud:2000un} on all our data and average
over the discrete rotational and parity symmetries of $S_L(p;a)$
to increase the statistics.

\subsection{The lattice quark propagator at tree-level}
For the sake of easier notation we will suppress the $a$ dependence
of the lattice quark propagator and write $Z_L(p)$ and $M_L(p)$ as 
functions of $p$ rather than $p^2$ in the following discussion.

The lattice quark propagator at tree-level $S_L^{(0)}(p)$ differs from the continuum 
case, \Eq{treelevelcont}, due to discretization artifacts,
\be\label{treelevellat}
	S_L^{(0)}(p) = \left(ia\kslash +aM_L^{(0)}(p)\right)^{-1}.
\ee
The dressing function $A_L^{(0)}(p)$ is by construction equal to one
at tree-level (at least without tree-level improvement) 
and thus the function $B_L^{(0)}(p)$ equals 
at tree-level the mass function $M_L^{(0)}(p)$.

We extract the CI lattice momentum $ak(p)$ from the tree-level propagator 
on the lattice and depict it
in \Fig{momenta}.
The result is consistent with the analytically derived expression for
the $\DCI$ momenta given in Appendix \ref{sec:analyt}.
\begin{figure}[h]
	\centering
	\includegraphics[width=1.0\columnwidth]{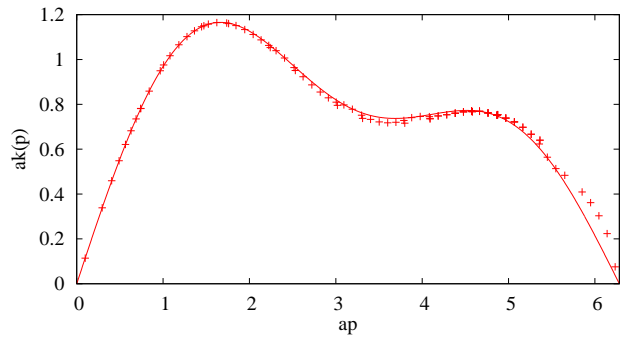}
	\caption{CI lattice momentum $ak(p)$ extracted from the tree-level propagator (crosses)
	compared to the analytical expression (full line) given in Appendix \ref{sec:analyt}.
	}\label{momenta}
\end{figure}

\begin{figure}[h]
	\centering
	\includegraphics[width=1.0\columnwidth]{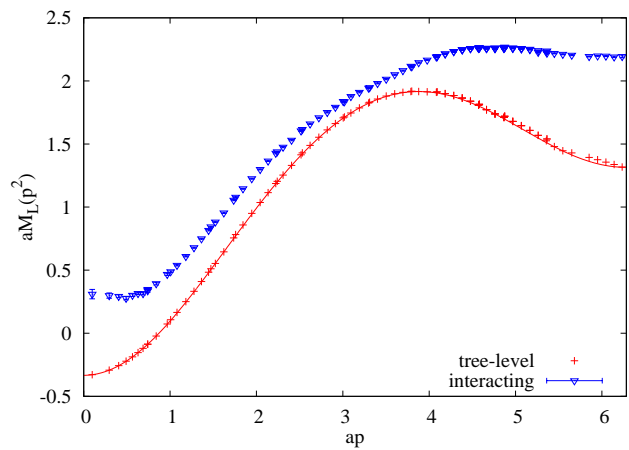}
	\caption{The lattice quark propagator mass function 
	at tree-level (red crosses and full line) and in the unimproved full interacting 
	case (blue triangles) without tree-level correction.
	The tree-level results comprise a lattice extraction from the tree-level 
	$\DCI$ (red crosses)
	and a plot of the analytical expression of the mass function (red line) given in Appendix \ref{sec:analyt}. 
	}\label{M_tree-level}
\end{figure}

The tree-level mass function $aM_L^{(0)}(p)$ which in the continuum  equals the bare mass $m$,
is shown in \Fig{M_tree-level} (red crosses), again together with the 
corresponding analytical expression.
We find that $aM_L^{(0)}(p)$ has a zero-crossing and $aM_L^{(0)}(0)\approx -0.333$.
The latter value 
is trivially equal to the sum of 
all coefficients of \Eq{DCIansatz} that come with a unit matrix in Dirac space
\be
	\sum_is_i+m_0
\ee
whereby the bare mass parameter is $m_0=-0.077$
and the non-zero $s_i$ are listed in Appendix \ref{sec:analyt}.

\subsection{The interacting propagator}
\begin{figure}[h]
	\centering
	\includegraphics[width=1.0\columnwidth]{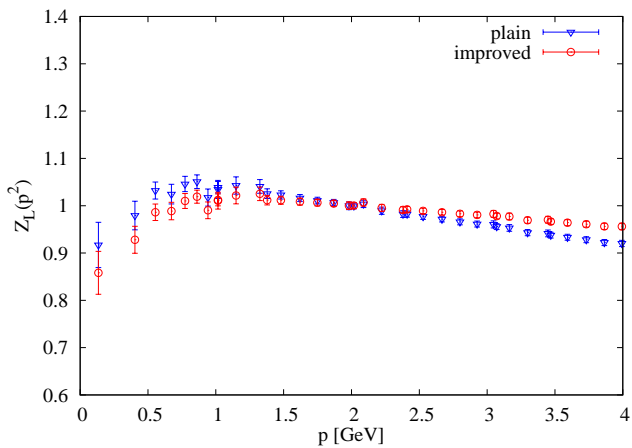}
	\caption{The wave-function renormalization function $Z_L(p)$ of the CI
	quark propagator: unimproved and without tree-level correction (blue 
	triangles) and with tree-level improvement and tree-level correction
	(red circles).
	The renormalization point is set at $\mu=\unit[2]{GeV}$.
	}\label{Z_unimproved}
\end{figure}

We expect the interacting propagator to have a similar form to the continuum case \Eq{fullcont},
hence we write
\be\label{interactlat}
	S_L(p) = Z_L(p) \left(ia\kslash +aM_L(p)\right)^{-1}.
\ee
The functions  $aM_L(p)$ and $Z_L(p)$ extracted from the lattice
Monte Carlo simulation 
are shown in 
\Fig{M_tree-level} and \Fig{Z_unimproved} (blue triangles),
respectively.
The shape of $aM_L(p)$ is similar to $aM_L^{(0)}(p)$ and
also $Z_L(p)$ strongly deviates from the expected monotonically growing behavior,
thus is clearly altered by 
discretization errors.

To get a handle on the lattice artifacts, i.e., to retain the
shapes of the wave-function renormalization function and the
mass function familiar from earlier lattice works as well as from 
Dyson-Schwinger equation studies \cite{Alkofer:2000wg}, we discuss
improvement and tree-level correction in the forthcoming subsections.

\subsection{Improvement}
The Symanzik improvement program \cite{Symanzik:1983dc}
offers a systematic way to reduce the errors of the fermionic action to $\bigO(a^2)$.
All terms that have the correct dimensionality and the symmetries of the QCD fermionic Lagrangian
must be included into the action:
\begin{align}
	L_1(x) &= \bar\psi\sigma_{\mu\nu}F_{\mu\nu}\psi, \\
	L_2(x) &= \bar\psi\overrightarrow{D}_\mu\overrightarrow{D}_\mu\psi
		+\bar\psi\overleftarrow{D}_\mu\overleftarrow{D}_\mu\psi, \\
	L_3(x) &= m\tr{F_{\mu\nu}F_{\mu\nu}}, \\
	L_4(x) &= m\left(\bar\psi\gamma_\mu\overrightarrow{D}_\mu\psi
		-\bar\psi\gamma_\mu\overleftarrow{D}_\mu\psi\right), \\
	L_5(x) &= m^2\bar\psi\psi.
\end{align} 
The terms $L_3$ and $L_5$ can be accounted for by a redefinition of the bare parameters $m$ and $g$.
$L_2$ and $L_4$ are only needed for off-shell quantities  
like hadronic matrix elements or the quark propagator \cite{Luscher:1996sc}.
Thus for on-shell quantities it is sufficient to take the
clover term \cite{Sheikholeslami:1985ij} (which corresponds to $L_1$) into account.

Note that whereas exact GW fermions are automatically $\bigO(a)$ improved, 
the CI operator fulfills the GW equation only approximately and thus
the clover term is included in the CI action.

Since the quark propagator is an off-shell quantity we would like 
to include the terms $L_2$ and $L_4$ as well.
In \cite{Heatlie:1990kg} it is shown that at tree-level $L_2$ and $L_4$ can be eliminated
by a transformation of the fermion fields according to
\begin{align}
	\psi &\to \left(1+\frac{a}{4}m\right)\left(1-\frac{a}{4}\gamma_\mu\overrightarrow{D}_\mu\right)\psi, \\
	\bar\psi &\to \left(1+\frac{a}{4}m\right)\bar\psi\left(1+\frac{a}{4}\gamma_\mu\overleftarrow{D}_\mu\right).
\end{align}

Improvement beyond tree-level requires tuning of the coefficients
of the fermion field transformations \cite{Dawson:1997gp}
which we do not attempt.
Hence we adopt the above fermion field transformations under which the quark propagator
turns into \cite{Skullerud:2000un, Skullerud:2001aw}
\be\label{SI}
	S_I(x,y)\equiv \left< (1+am)S(x,y;U)-\frac{a}{2}\delta(x,y)\right>
\ee 
where the index $I$ denotes improvement. In \Eq{SI},
$S(x,y;U)$ is obtained by inverting the $\DCI$ operator on each configuration
and the brackets
denote Monte Carlo integration over the gauge fields $U$.

All results that follow have been tree-level improved according to
the above prescription.

\subsection{Tree-level correction}
In order to blank out the lattice artifacts which are already present
at tree-level, we now focus on the derivation of the interacting propagator
from its tree-level form.

For the renormalization function $Z_L(p)$
we adopt a multiplicative
tree-level correction
\be
	Z_L(p) \to \frac{Z_L(p)}{Z_L^{(0)}(p)}.
\ee
As can be seen in \Fig{Z_unimproved} (red circles), this procedure together with the
tree-level improvement from the previous subsection flattens $Z_L(p)$, hence
reduces the dominant lattice artifacts.
However, the fact that the function is still not monotonically growing
indicates that the improvement
coefficients are not sufficiently adjusted to remove all $\bigO(a)$ errors
when simply picking their tree-level
values.

In order to apply a multiplicative tree-level correction 
to the mass function of the form
\be
	aM_L(p) \to \frac{amM_L(p)}{M_L^{(0)}(p)}
\ee
we have to carry out an additive mass renormalization of the tree-level function $B_L^{(0)}(p)$ 
in order to avoid divergences, i.e., 
\be
	aB_L^{(0)}(p)\to aB_L^{(0)}(p) + am_\textrm{add}
\ee
where $am_\textrm{add}$ is chosen such that $B_L^{(0)}(0)=m$, like in the continuum,
thus
\be
	am_\textrm{add} = am-aB_L^{(0)}(0) \approx 0.344.
\ee
As a result, the multiplicative tree-level correction 
for the mass function is
\be 
	aM_L(p) \to \frac{amM_L(p)A_L^{(0)}(p)}{B_L^{(0)}(p)+m_\textrm{add}}.
\ee

Alternatively, we may adopt an
hybrid tree-level correction which is based on the ideas developed in Ref. \cite{Skullerud:2001aw}:
if $p<p'$, then perform 
an additive tree-level correction
\be
	aM_L(p) \to aM_L(p) - \frac{aB_L^{(0)}(p) + am_\textrm{add}}{A_L^{(0)}(p)}
\ee
and for momenta larger than $p'$ apply a multiplicative tree-level correction
\be 
	aM_L(p) \to \frac{amM_L(p)A_L^{(0)}(p)}{B_L^{(0)}(p)}.
\ee
The momentum parameter $p'$ should be adjusted thereby such that $M_L(p)$ is continuous
and smooth at $p=p'$ which we found to be the case for  $p'=\unit[1.5]{GeV}$.

\begin{figure}[h]
	\centering
	\includegraphics[width=1.0\columnwidth]{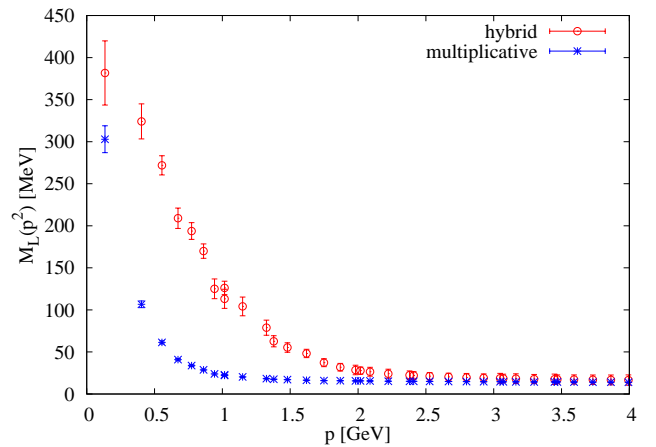}
	\caption{The CI quark propagator mass function $M_L(p)$ after
	improvement and application of a pure multiplicative (blue crosses) 
	and an hybrid (red circles) tree-level correction procedure.
	}\label{M_hybrid}
\end{figure}

Both possibilities of tree-level correction
for the mass function $M_L(p)$ are plotted
in \Fig{M_hybrid}.
We observe that the pure multiplicative correction (blue crosses) results
in an infrared enhanced function, enhancement occurring from \unit[1.25]{GeV} on
downwards and appearing to be rather steep.
The hybrid scheme (red circles), on the other hand, exhibits a wider range of IR mass generation (from \unit[2.5]{GeV} on
downwards), gives a higher
IR mass and yields flattening of the mass function in the deep IR. 
The hybrid scheme allows for an earlier mass generation due to the fact that
the multiplicative correction therein (for $p\geq p'$) does not require an
additive mass renormalization since the zero-crossing of the tree-level function
is handled by the additive tree-level correction ($p<p'$).

When comparing these results to lattice quark propagator studies from a different 
fermionic action,
for example to the (quenched) overlap quark propagator
\cite{Bonnet:2002ih, Zhang:2003faa, Zhang:2004gv, Kamleh:2004aw, Kamleh:2007ud, Bowman:2005zi, Bowman:2004xi},
we find better agreement for the hybrid scheme.
It has to be stressed however that the parameter $p'$ introduces a small 
arbitrariness to the procedure whereas the simpler pure multiplicative scheme
provides a straightforward comparison of the interacting mass function to its
tree-level counterpart while still yielding \emph{qualitatively}
the correct physics.
Consequently, for the next section we adopt the simpler multiplicative scheme 
for the analysis
of the effects of Dirac low-mode removal on the quark propagator mass function in 
order to avoid possible systematic errors related to the tuning of $p'$.

\section{Restoration of chiral symmetry}\label{sec:restoration}
The lowest Dirac eigenmodes are known to play a crucial role for dynamical chiral
symmetry breaking as stated by the Banks--Casher relation \cite{Banks:1979yr}.
The latter relates the chiral condensate to the density of the
smallest nonzero Dirac eigenmodes.
As a consequence, when removing the Dirac eigenmodes near the origin from the theory,
the chiral condensate vanishes and chiral symmetry becomes 
``artificially restored'' \cite{Lang:2011vw}.

The aim of the current work is
to analyze the effects of artificial chiral restoration on the 
dressing functions of the quark propagator.
Consider the Hermitian Dirac operator $D_5\equiv \gamma_5D$
which is normal and thus has real eigenvalues $\mu_i$. $D$ can be written in terms of the spectral representation of $D_5$,
\be\label{gafDspectral}
	D = \sum_{i=1}^N \,\mu_i\,\gamma_5\,\ket{v_i}\bra{v_i}.
\ee
We split the quark propagator $S=D^{-1}$ into a low-mode part (lm) and a reduced part (red), e.g., using the eigenvalues and eigenvectors of $D_5$,
\begin{align}
	S &= \sum_{i\leq k}\, \mu_i^{-1} \,\ket{v_i}\bra{v_i}\,\gaf 
	  + \sum_{i>k}\,\mu_i^{-1}\,\ket{v_i}\bra{v_i}\,\gaf\\
	  &\equiv S_\LM{k} + S_\RD{k}.
\end{align}
Hence we can obtain the reduced part of the propagator by subtracting the low-mode part from the full propagator
\be\label{SRD}
	S_\RD{k} = S - S_\LM{k}.
\ee

\begin{figure}[h]
	\centering
	\includegraphics[width=1.0\columnwidth]{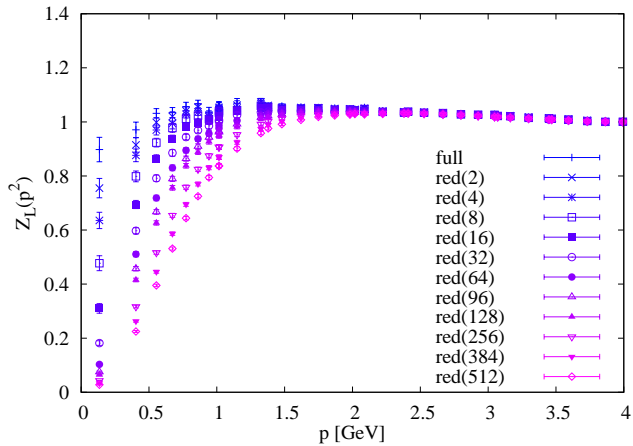}\\
	\caption{The quark wave-function renormalization function $Z_L(p)$
	under Dirac eigenmode removal for different reduction levels $k$.
	The renormalization point is set at $\mu=\unit[4]{GeV}$.
	}\label{Z_red}
\end{figure}

\begin{figure}[h]
	\centering
	\includegraphics[width=1.0\columnwidth]{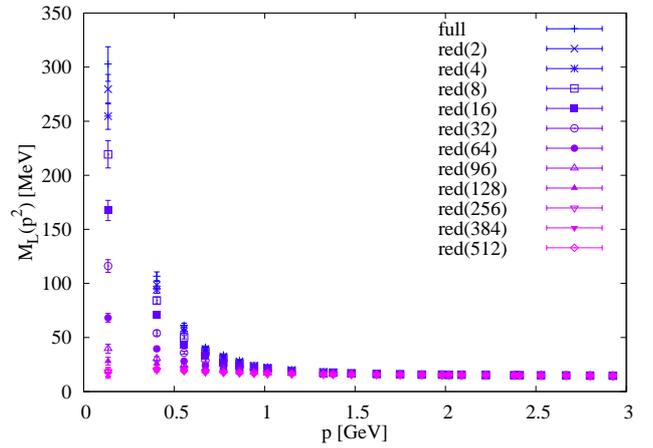}
	\caption{The quark mass function $M_L(p)$
	under Dirac eigenmode removal for different reduction levels $k$.
	}\label{M_red}
\end{figure}

We calculate the quark wave-function renormalization function $Z_L(p)$ and the quark
mass function $M_L(p)$ from the reduced propagators of \Eq{SRD} with varying reduction levels
$k=2-512$.
We tree-level improve the modified propagators and apply the
multiplicative tree-level correction scheme, cf. \Sec{sec:CIquark}.
The dressing functions from reduced propagators are presented in \Fig{Z_red} and \Fig{M_red}.

Figure \ref{Z_red} reveals amplification of IR suppression of $Z_L(p)$
when subtracting 
Dirac low-modes whereas the range from medium to high momenta is not
altered at all.
The mass function $M_L(p)$, \Fig{M_red}, demonstrates a similar behavior: it gets suppressed
in the IR when removing more and more eigenmodes until the dynamic
generation of mass completely ceases at truncation stage $k\approx 128$.

\begin{figure}[h]
	\centering
	\includegraphics[width=1.0\columnwidth]{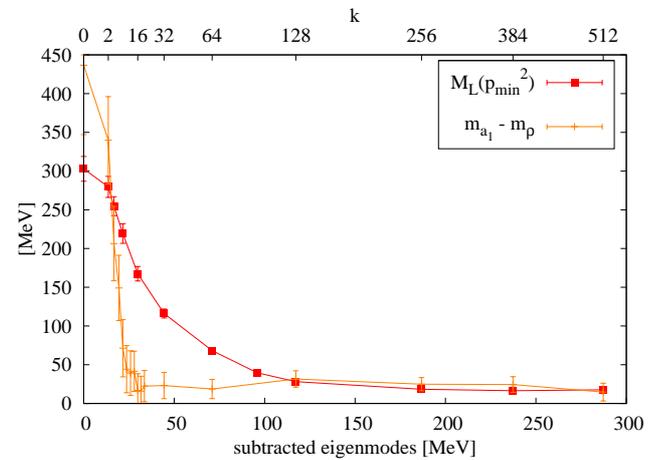}
	\caption{The infrared mass $M_L(p_{\textrm{min}}^2)$ of the reduced CI quark propagator
	as a function of the reduction level
	compared to the mass splitting between the $\rho$ and the $a_1$
	from Ref. \cite{Lang:2011vw}. 
	The upper abscissa shows the truncation level $k$ and the lower abscissa gives
	the corresponding energy scale, the relation between the two is obtained by
	integrating the histograms of the $D_5$ eigenvalues. 
	}\label{Mpmin}
\end{figure}

In \Fig{Mpmin} we compare the deep IR mass of the CI quark propagator
from $M_L(p_{\textrm{min}}^2)$, at the smallest available momentum 
$p_{\textrm{min}}=\unit[0.1345]{GeV}$, as a function of the reduction level
to the mass splitting of the vector meson $\rho$ and its chiral 
partner the axial vector current $a_1$, taken from Ref. \cite{Lang:2011vw}.
Note that the reduction level $k$ can be translated to an energy scale which
is given in the lower abscissa of the figure and was derived in \cite{Lang:2011vw}
by integrating the histograms of the eigenvalues. 

The mass splitting between the $\rho$ and
the $a_1$ rapidly drops down and reaches a plateau after subtracting 
about 16 eigenmodes; it does not go down to zero which can most likely
be attributed to the small explicit chiral symmetry breaking by the nonvanishing
quark mass.
In contrast, the dynamically generated mass of the quark propagator,
$M_L(p_{\textrm{min}}^2)$, decreases slower and reaches its plateau
only after subtracting more than 128 Dirac eigenmodes.

\section{Conclusions}\label{sec:conclusions}

The wave-function renormalization function $Z(p^2)$ and the mass function $M(p^2)$
from the CI quark propagator have been analyzed on configurations
with two light degenerate CI quark flavors.
It has been demonstrated that the combination of tree-level improvement and a multiplicative
or hybrid tree-level correction scheme drastically reduce the
dominant lattice artifacts.

Removing the lowest Dirac eigenmodes out of the quark propagator strongly suppresses 
the wave-function renormalization function in the IR and completely 
dissolves the dynamically generated mass displayed by $M(p^2)$. 
Under Dirac low-mode removal,
the mass function is found to reveal a smoother
transition towards
chiral restoration
than the splitting of the vector
and axial vector currents.

\begin{acknowledgments}
I thank Joseph Day for proof reading of the manuscript 
and I am very grateful to L.Ya.~Glozman for valuable discussions.
Special thanks are addressed to C.B.~Lang 
for helpful advice and clarifying discussions.
The calculations have been performed on clusters at ZID at the University of Graz. 
Support by the Austrian Science Fund (FWF) DK W1203-N16
and the Research Executive Agency (REA) of the European Union under Grant Agreement 
PITN-GA-2009-238353 (ITN STRONGnet) is gratefully acknowledged.\\
\end{acknowledgments}


\appendix

\section{Gauge fixing on the GPU}\label{sec:GPU}

In the current Appendix
we discuss how the process of lattice gauge fixing with the overrelaxation 
algorithm can be accelerated by using NVIDIA\symbR's CUDA\symbTM (Compute Unified Device 
Architecture) programming environment for GPUs. 
We compare the performance of the overrelaxation
algorithm on one  GPU (NVIDIA GeForce\symbR GTX 580) with conventional calculations on the CPU and apply techniques to 
relax the bandwidth restrictions of the GPU.

In the recent years, many groups in the lattice QCD community have taken advantage
of the cost effective opportunity to adopt GPUs
for high-performance lattice QCD computations.
Whereas the pioneering work of GPU calculations in lattice QCD was reported in \cite{Egri:2006zm},
some more recent examples are given by 
\cite{Babich:2010mu, Clark:2009wm, Winter:2011dh, Lujan:2011ue, Cardoso:2011xu, Jang:2011cn, Walk:2010ut, Cardoso:2010di}.

\subsection{Gauge fixing via (over)relaxation}

The underlying idea of the relaxation algorithm \cite{Mandula:1987rh} is 
a local optimization of the gauge functional $F_g[U]$,
i.e., for all $x$ the maximum of $\Re\tr{g(x)K(x)}$ is wanted,
where
\begin{equation}\label{Kx}
	K(x)\equiv \sum_\mu\Big( U_\mu(x) g(x+\hat\mu)^\dagger 
		+ U_\mu(x-\hat\mu)^\dagger g(x-\hat\mu)^\dagger\Big).
\end{equation}
The solution of the aforementioned subtask is given, in the case of the gauge group $\mathrm{SU}(2)$, by
\be\label{localsolution}  
g(x) = \frac{K(x)^\dagger}{\sqrt{\det{K(x)^\dagger}}} 
\ee
and for $\mathrm{SU}(3)$ one iteratively operates in the three subgroups of $\mathrm{SU}(2)$.
From equations \eqref{Kx} and \eqref{localsolution} it is evident that 
one can optimize all sites in 
each of the black and white checker sub-lattices simultaneously.

In order to reduce the critical slowing down of the relaxation algorithm on large
lattices, the authors of \cite{Mandula:1990vs} suggested to apply an \emph{over}relaxation algorithm which replaces
the gauge transformation $g(x)$ by $g^\omega(x)$ in each step of the iteration. This
method has widely been studied and the
value of $\omega$ was found to be well adapted at around 1.7, see \cite{Giusti:2001xf}
and references therein.

\subsection{Lattice QCD on the GPU}

Since CUDA supports only lattices up to three dimensions natively, 
one single index that runs over all four dimensions
of the  space--time lattice is used.
We assign each lattice site to a separate thread and start 32 threads per multiprocessor.
Better occupancy would be achieved with more threads per multiprocessor but  we are
restricted by the 48 KB of L1 cache.

A function which is called from the host system and which performs calculations on the GPU is called
a kernel. We implemented two kernels, one which checks the current value of the gauge fixing
functional, \Eq{gaugefunc}, and the gauge precision, \Eq{gaugeprec}, after every 100th iteration step and a second 
which does the actual work,
i.e., which performs an overrelaxation step. The latter is invoked for 
black and white
lattice sites consecutively.

\subsection{Optimization}

The GPU can read data from global device memory in an efficient way only if the data is accurately 
coalesced; that means the largest memory throughput is achieved when consecutive threads
read from consecutive memory addresses. 
In order to do so we rearrange the gauge field into two blocks, one for even and one
for odd lattice sites. Moreover, for the same sake of memory coalescing, we choose the site index running 
faster than color and Dirac indices.

Applying the overrelaxation algorithm to one lattice site requires 
2253 floating point operations and
we have to read and write eight $\mathrm{SU}(3)$ matrices for every site; 
thus the required data transfer in single precision is 1152 bytes per site. Comparing the ratio 
data transfer per floating point operation, $1152/2253\approx 0.51$,
with the theoretical peak performance of the GTX 580, $192/1581\approx 0.12$, we clearly see that
we are solely constrained by memory bandwidth and not by the maximum number of arithmetic 
instructions. 

In order to reduce memory traffic we make use of the unitarity of $\mathrm{SU}(3)$ matrices
and reconstruct the third line of each matrix on the fly when needed instead of storing it \cite{Babich:2010mu}.
A minimal 8 parameter reconstruction turned out to be numerically not stable
enough for our purpose since we not only read but also write the modified gauge fields 
in each step of the iteration.
For more details see \cite{Babich:2010mu, Clark:2009wm}
and references therein.

\subsection{Performance}
We generated quenched $\mathrm{SU}(3)$ configurations with the heat-bath algorithm with 
$24^3$ spatial lattice sites
and a varying temporal extent from $4$ to $128$.
On these configurations
we compare the performance of our GPU kernels on the GTX 580 in single and
double precision to the performance of the equivalent code (with a data alignment more
appropriate for the CPU) on one
core of the Intel\symbR Core\symbTM i7-950 (``Nehalem Bloomfield'') processor run at \unit[3.06]{GHz},
whereby the CPU code is optimized through SSE4 instructions generated
by the compiler.
\footnote{Using the Intel compiler (12.0.0) with the compiler flag SSE4.2.}

\begin{figure}[h]
	\centering
	\includegraphics[width=0.49\columnwidth]{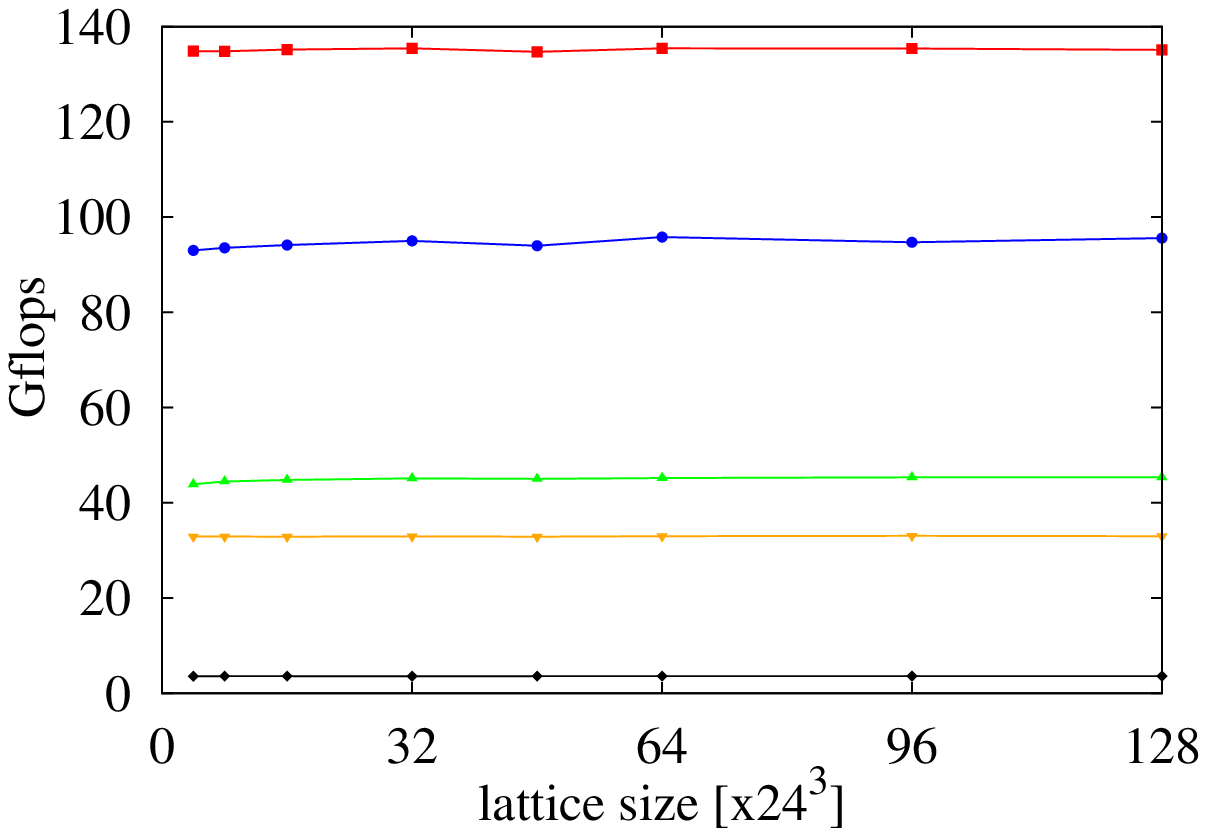}
	\includegraphics[width=0.49\columnwidth]{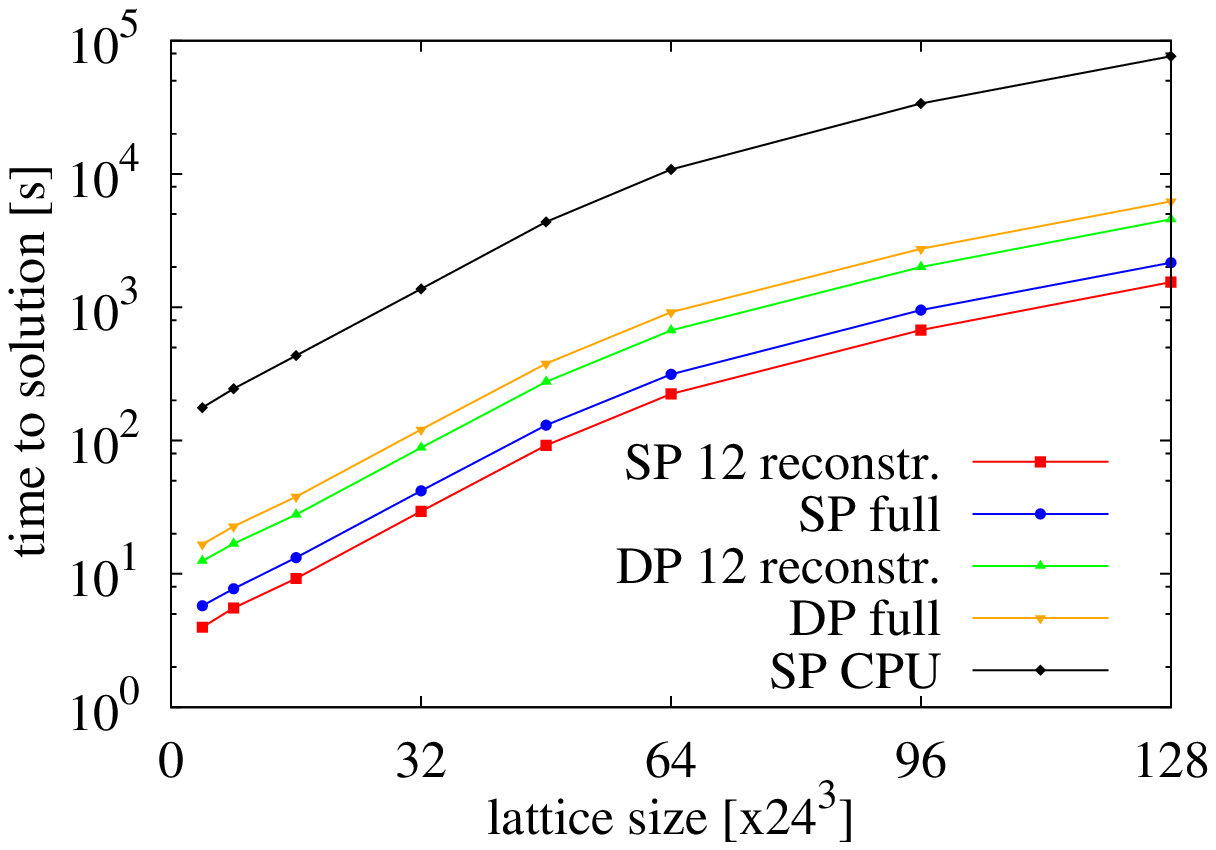}
	\caption{Performance of the overrelaxation algorithm for fixing the gauge on the 
	GPU (NVIDIA GeForce GTX 580)
	in single precision (SP) and double precision (DP) with and without the 12 parameter reconstruction
	for $\SU{3}$ matrices described in the text, compared to the performance on one core of the 
	Intel Core i7-950 processor (CPU) in single precision.
	On the l.h.s.~shown in terms of Gflops and on the r.h.s.~in terms of time (seconds) to solution.
	}\label{gpu}
\end{figure}

The results of the performance test are given in \Fig{gpu}:
in the l.h.s.~plot we show the performance of the algorithm using a 12 parameter 
reconstruction and a full 18 number representation in single and double precision
together with the performance of the same code run on the CPU in single precision. 
We achieve a maximum 
performance of 135 Gflops (independent of the lattice size) for the case of the 12 parameter 
reconstruction in single precision.
On the r.h.s.~of \Fig{gpu} we present a summary of the time needed to fix the gauge for the
various settings up to the test accuracy of $\theta< 10^{-6}$.
Overall, we find that for the task of gauge fixing with the overrelaxation algorithm
the computational power of one GPU is equivalent to approximately 40 CPU cores
(under the assumption of ideal scaling).\\

\section{Analytical expressions for the tree-level CI Dirac operator}\label{sec:analyt}
At tree-level, the tensor, axialvector and pseudoscalar terms of \Eq{DCIansatz} vanish identically and only scalar and vector terms remain \cite{Gattringer:2008vj}.
When transformed to momentum space one obtains the following analytical 
expressions for the latter two:
the scalar part, i.e., the tree-level mass function which is
plotted in \Fig{M_tree-level} is given by
\begin{align*}
		&M^{(0)}_L(p) = s_1 + 48 s_{13}\\
		&+ (2 s_2 + 12 s_8) (\cos(p_0) + \cos(p_1) + \cos(p_2) + \cos(p_3))\\
		&+ (8 s_3 + 64 s_{11}) (\cos(p_0)  \cos(p_1) + \cos(p_0)  \cos(p_2) \\
		&+ \cos(p_0)  \cos(p_3) + \cos(p_1)  \cos(p_2) + \cos(p_1)  \cos(p_3) \\
		&+ \cos(p_2)  \cos(p_3))\\
		&+ 48 s_5 (\cos(p_0)  \cos(p_1)  \cos(p_2) + \cos(p_0)  \cos(p_1)  \cos(p_3)\\
		&+ \cos(p_0)  \cos(p_2)  \cos(p_3) + \cos(p_1)  \cos(p_2)  \cos(p_3))\\
		&+ 8 s_6 (\cos(p_0)  \cos(2 p_1) + \cos(p_0)  \cos(2 p_2)\\
		&+ \cos(p_0)  \cos(2 p_3) + \cos(p_1)  \cos(2 p_2)\\
		&+ \cos(p_1)  \cos(2 p_3) + \cos(p_2)  \cos(2 p_3)\\
		&+ \cos(2 p_0)  \cos(p_1) + \cos(2 p_0)  \cos(p_2)\\
		&+ \cos(2 p_0)  \cos(p_3) + \cos(2 p_1)  \cos(p_2)\\
		&+ \cos(2 p_1)  \cos(p_3) + \cos(2 p_2)  \cos(p_3))\\
		&+ 384 s_{10}  \cos(p_0)  \cos(p_1)  \cos(p_2)  \cos(p_3)\\
		&+ m_0,
\end{align*}
where the relevant coefficients are listed in \Tab{tab:CIparams}.
In the same manner, the analytical expressions of the
lattice momenta $k_\mu(p_\mu)$ from \Fig{momenta} read
\begin{align*}
		k_0 &= 2 v_1  \sin(p_0) + 8 v_2  \sin(p_0) (\cos(p_1) + \cos(p_2) + \cos(p_3))\\
		&+ (32 v_4 + 16 v_5)  \sin(p_0) (\cos(p_1)  \cos(p_2) + \cos(p_1)  \cos(p_3)\\
		&+ \cos(p_2)  \cos(p_3)),\\
		k_1 &= 2 v_1  \sin(p_1) + 8 v_2  \sin(p_1) (\cos(p_0) + \cos(p_2) + \cos(p_3))\\
		&+ (32 v_4 + 16 v_5)  \sin(p_1) (\cos(p_0)  \cos(p_2) + \cos(p_0)  \cos(p_3)\\
		&+ \cos(p_2)  \cos(p_3)),\\
		k_2 &= 2 v_1  \sin(p_2) + 8 v_2  \sin(p_2) (\cos(p_0) + \cos(p_1) + \cos(p_3))\\
		&+ (32 v_4 + 16 v_5)  \sin(p_2) (\cos(p_0)  \cos(p_1) + \cos(p_0)  \cos(p_3)\\
		&+ \cos(p_1)  \cos(p_3)),\\
		k_3 &= 2 v_1  \sin(p_3) + 8 v_2  \sin(p_3) (\cos(p_0) + \cos(p_1) + \cos(p_2))\\
		&+ (32 v_4 + 16 v_5)  \sin(p_3) (\cos(p_0)  \cos(p_1) + \cos(p_0)  \cos(p_2)\\
		&+ \cos(p_1)  \cos(p_2)).
\end{align*}
The wave-function renormalization function 
is equal to one at tree-level by construction.

\begin{table}[h]
	\center
	\begin{tabular}{lc}
	\hline\hline
	 $s_{1}$ & $ 0.1481599252\times 10^{1}$\\ 
	 $s_{2}$ & $-0.5218251439\times 10^{-1}$\\ 
	 $s_{3}$ & $-0.1473643847\times 10^{-1}$\\ 
	 $s_{5}$ & $-0.2186103421\times 10^{-2}$\\ 
	 $s_{6}$ & $ 0.2133989696\times 10^{-2}$\\ 
	 $s_{8}$ & $-0.3997001821\times 10^{-2}$\\ 
	 $s_{10}$ & $-0.4951673735\times 10^{-3}$\\
	 $s_{11}$ & $-0.9836500799\times 10^{-3}$\\
	 $s_{13}$ & $ 0.7529838581\times 10^{-2}$\\
	 $v_{1}$ & $ 0.1972229309\times 10^{0}$\\ 
	 $v_{2}$ & $ 0.8252157565\times 10^{-2}$\\ 
	 $v_{4}$ & $ 0.5113056314\times 10^{-2}$\\ 
	 $v_{5}$ & $ 0.1736609425\times 10^{-2}$\\ 
	 $m_0$ & $-0.077$ \\\hline\hline
	\end{tabular}
	\caption{The relevant $\DCI$ coefficients.
	For a complete description see \cite{Gattringer:2008vj}.}
	\label{tab:CIparams}
\end{table}

\bibliographystyle{apsrev4-1}

\begin{thebibliography}{52}%
\makeatletter
\providecommand \@ifxundefined [1]{%
 \@ifx{#1\undefined}
}%
\providecommand \@ifnum [1]{%
 \ifnum #1\expandafter \@firstoftwo
 \else \expandafter \@secondoftwo
 \fi
}%
\providecommand \@ifx [1]{%
 \ifx #1\expandafter \@firstoftwo
 \else \expandafter \@secondoftwo
 \fi
}%
\providecommand \natexlab [1]{#1}%
\providecommand \enquote  [1]{``#1''}%
\providecommand \bibnamefont  [1]{#1}%
\providecommand \bibfnamefont [1]{#1}%
\providecommand \citenamefont [1]{#1}%
\providecommand \href@noop [0]{\@secondoftwo}%
\providecommand \href [0]{\begingroup \@sanitize@url \@href}%
\providecommand \@href[1]{\@@startlink{#1}\@@href}%
\providecommand \@@href[1]{\endgroup#1\@@endlink}%
\providecommand \@sanitize@url [0]{\catcode `\\12\catcode `\$12\catcode
  `\&12\catcode `\#12\catcode `\^12\catcode `\_12\catcode `\%12\relax}%
\providecommand \@@startlink[1]{}%
\providecommand \@@endlink[0]{}%
\providecommand \url  [0]{\begingroup\@sanitize@url \@url }%
\providecommand \@url [1]{\endgroup\@href {#1}{\urlprefix }}%
\providecommand \urlprefix  [0]{URL }%
\providecommand \Eprint [0]{\href }%
\providecommand \doibase [0]{http://dx.doi.org/}%
\providecommand \selectlanguage [0]{\@gobble}%
\providecommand \bibinfo  [0]{\@secondoftwo}%
\providecommand \bibfield  [0]{\@secondoftwo}%
\providecommand \translation [1]{[#1]}%
\providecommand \BibitemOpen [0]{}%
\providecommand \bibitemStop [0]{}%
\providecommand \bibitemNoStop [0]{.\EOS\space}%
\providecommand \EOS [0]{\spacefactor3000\relax}%
\providecommand \BibitemShut  [1]{\csname bibitem#1\endcsname}%
\let\auto@bib@innerbib\@empty
\bibitem [{\citenamefont {Skullerud}\ and\ \citenamefont
  {Williams}(2001)}]{Skullerud:2000un}%
  \BibitemOpen
  \bibfield  {author} {\bibinfo {author} {\bibfnamefont {J.~I.}\ \bibnamefont
  {Skullerud}}\ and\ \bibinfo {author} {\bibfnamefont {A.~G.}\ \bibnamefont
  {Williams}},\ }\href {\doibase 10.1103/PhysRevD.63.054508} {\bibfield
  {journal} {\bibinfo  {journal} {Phys.Rev.}\ }\textbf {\bibinfo {volume}
  {D63}},\ \bibinfo {pages} {054508} (\bibinfo {year} {2001})},\ \Eprint
  {http://arxiv.org/abs/hep-lat/0007028} {arXiv:hep-lat/0007028 [hep-lat]}
  \BibitemShut {NoStop}%
\bibitem [{\citenamefont {Skullerud}\ \emph {et~al.}(2001)\citenamefont
  {Skullerud}, \citenamefont {Leinweber},\ and\ \citenamefont
  {Williams}}]{Skullerud:2001aw}%
  \BibitemOpen
  \bibfield  {author} {\bibinfo {author} {\bibfnamefont {J.}~\bibnamefont
  {Skullerud}}, \bibinfo {author} {\bibfnamefont {D.~B.}\ \bibnamefont
  {Leinweber}}, \ and\ \bibinfo {author} {\bibfnamefont {A.~G.}\ \bibnamefont
  {Williams}},\ }\href {\doibase 10.1103/PhysRevD.64.074508} {\bibfield
  {journal} {\bibinfo  {journal} {Phys.Rev.}\ }\textbf {\bibinfo {volume}
  {D64}},\ \bibinfo {pages} {074508} (\bibinfo {year} {2001})},\ \Eprint
  {http://arxiv.org/abs/hep-lat/0102013} {arXiv:hep-lat/0102013 [hep-lat]}
  \BibitemShut {NoStop}%
\bibitem [{\citenamefont {Kogut}\ and\ \citenamefont
  {Susskind}(1975)}]{Kogut:1974ag}%
  \BibitemOpen
  \bibfield  {author} {\bibinfo {author} {\bibfnamefont {J.~B.}\ \bibnamefont
  {Kogut}}\ and\ \bibinfo {author} {\bibfnamefont {L.}~\bibnamefont
  {Susskind}},\ }\href {\doibase 10.1103/PhysRevD.11.395} {\bibfield  {journal}
  {\bibinfo  {journal} {Phys.Rev.}\ }\textbf {\bibinfo {volume} {D11}},\
  \bibinfo {pages} {395} (\bibinfo {year} {1975})}\BibitemShut {NoStop}%
\bibitem [{\citenamefont {Orginos}\ \emph {et~al.}(1999)\citenamefont
  {Orginos}, \citenamefont {Toussaint},\ and\ \citenamefont
  {Sugar}}]{Orginos:1999cr}%
  \BibitemOpen
  \bibfield  {author} {\bibinfo {author} {\bibfnamefont {K.}~\bibnamefont
  {Orginos}}, \bibinfo {author} {\bibfnamefont {D.}~\bibnamefont {Toussaint}},
  \ and\ \bibinfo {author} {\bibfnamefont {R.}~\bibnamefont {Sugar}} (\bibinfo
  {collaboration} {MILC Collaboration}),\ }\href {\doibase
  10.1103/PhysRevD.60.054503} {\bibfield  {journal} {\bibinfo  {journal}
  {Phys.Rev.}\ }\textbf {\bibinfo {volume} {D60}},\ \bibinfo {pages} {054503}
  (\bibinfo {year} {1999})},\ \Eprint {http://arxiv.org/abs/hep-lat/9903032}
  {arXiv:hep-lat/9903032 [hep-lat]} \BibitemShut {NoStop}%
\bibitem [{\citenamefont {Bowman}\ \emph
  {et~al.}(2002{\natexlab{a}})\citenamefont {Bowman}, \citenamefont {Heller},\
  and\ \citenamefont {Williams}}]{Bowman:2002bm}%
  \BibitemOpen
  \bibfield  {author} {\bibinfo {author} {\bibfnamefont {P.~O.}\ \bibnamefont
  {Bowman}}, \bibinfo {author} {\bibfnamefont {U.~M.}\ \bibnamefont {Heller}},
  \ and\ \bibinfo {author} {\bibfnamefont {A.~G.}\ \bibnamefont {Williams}},\
  }\href {\doibase 10.1103/PhysRevD.66.014505} {\bibfield  {journal} {\bibinfo
  {journal} {Phys.Rev.}\ }\textbf {\bibinfo {volume} {D66}},\ \bibinfo {pages}
  {014505} (\bibinfo {year} {2002}{\natexlab{a}})},\ \Eprint
  {http://arxiv.org/abs/hep-lat/0203001} {arXiv:hep-lat/0203001 [hep-lat]}
  \BibitemShut {NoStop}%
\bibitem [{\citenamefont {Parappilly}\ \emph {et~al.}(2006)\citenamefont
  {Parappilly}, \citenamefont {Bowman}, \citenamefont {Heller}, \citenamefont
  {Leinweber}, \citenamefont {Williams} \emph {et~al.}}]{Parappilly:2005ei}%
  \BibitemOpen
  \bibfield  {author} {\bibinfo {author} {\bibfnamefont {M.~B.}\ \bibnamefont
  {Parappilly}}, \bibinfo {author} {\bibfnamefont {P.~O.}\ \bibnamefont
  {Bowman}}, \bibinfo {author} {\bibfnamefont {U.~M.}\ \bibnamefont {Heller}},
  \bibinfo {author} {\bibfnamefont {D.~B.}\ \bibnamefont {Leinweber}}, \bibinfo
  {author} {\bibfnamefont {A.~G.}\ \bibnamefont {Williams}},  \emph {et~al.},\
  }\href {\doibase 10.1103/PhysRevD.73.054504} {\bibfield  {journal} {\bibinfo
  {journal} {Phys.Rev.}\ }\textbf {\bibinfo {volume} {D73}},\ \bibinfo {pages}
  {054504} (\bibinfo {year} {2006})},\ \Eprint
  {http://arxiv.org/abs/hep-lat/0511007} {arXiv:hep-lat/0511007 [hep-lat]}
  \BibitemShut {NoStop}%
\bibitem [{\citenamefont {Bowman}\ \emph
  {et~al.}(2002{\natexlab{b}})\citenamefont {Bowman}, \citenamefont {Heller},\
  and\ \citenamefont {Williams}}]{Bowman:2001xh}%
  \BibitemOpen
  \bibfield  {author} {\bibinfo {author} {\bibfnamefont {P.~O.}\ \bibnamefont
  {Bowman}}, \bibinfo {author} {\bibfnamefont {U.~M.}\ \bibnamefont {Heller}},
  \ and\ \bibinfo {author} {\bibfnamefont {A.~G.}\ \bibnamefont {Williams}},\
  }\href {\doibase 10.1016/S0920-5632(01)01854-0} {\bibfield  {journal}
  {\bibinfo  {journal} {Nucl.Phys.Proc.Suppl.}\ }\textbf {\bibinfo {volume}
  {106}},\ \bibinfo {pages} {820} (\bibinfo {year} {2002}{\natexlab{b}})},\
  \Eprint {http://arxiv.org/abs/hep-lat/0110081} {arXiv:hep-lat/0110081
  [hep-lat]} \BibitemShut {NoStop}%
\bibitem [{\citenamefont {Bowman}\ \emph {et~al.}(2003)\citenamefont {Bowman},
  \citenamefont {Heller}, \citenamefont {Leinweber},\ and\ \citenamefont
  {Williams}}]{Bowman:2002kn}%
  \BibitemOpen
  \bibfield  {author} {\bibinfo {author} {\bibfnamefont {P.~O.}\ \bibnamefont
  {Bowman}}, \bibinfo {author} {\bibfnamefont {U.~M.}\ \bibnamefont {Heller}},
  \bibinfo {author} {\bibfnamefont {D.~B.}\ \bibnamefont {Leinweber}}, \ and\
  \bibinfo {author} {\bibfnamefont {A.~G.}\ \bibnamefont {Williams}},\ }\href
  {\doibase 10.1016/S0920-5632(03)01533-0} {\bibfield  {journal} {\bibinfo
  {journal} {Nucl.Phys.Proc.Suppl.}\ }\textbf {\bibinfo {volume} {119}},\
  \bibinfo {pages} {323} (\bibinfo {year} {2003})},\ \Eprint
  {http://arxiv.org/abs/hep-lat/0209129} {arXiv:hep-lat/0209129 [hep-lat]}
  \BibitemShut {NoStop}%
\bibitem [{\citenamefont {Bowman}\ \emph
  {et~al.}(2005{\natexlab{a}})\citenamefont {Bowman}, \citenamefont {Heller},
  \citenamefont {Leinweber}, \citenamefont {Parappilly}, \citenamefont
  {Williams} \emph {et~al.}}]{Bowman:2005vx}%
  \BibitemOpen
  \bibfield  {author} {\bibinfo {author} {\bibfnamefont {P.~O.}\ \bibnamefont
  {Bowman}}, \bibinfo {author} {\bibfnamefont {U.~M.}\ \bibnamefont {Heller}},
  \bibinfo {author} {\bibfnamefont {D.~B.}\ \bibnamefont {Leinweber}}, \bibinfo
  {author} {\bibfnamefont {M.~B.}\ \bibnamefont {Parappilly}}, \bibinfo
  {author} {\bibfnamefont {A.~G.}\ \bibnamefont {Williams}},  \emph {et~al.},\
  }\href {\doibase 10.1103/PhysRevD.71.054507} {\bibfield  {journal} {\bibinfo
  {journal} {Phys.Rev.}\ }\textbf {\bibinfo {volume} {D71}},\ \bibinfo {pages}
  {054507} (\bibinfo {year} {2005}{\natexlab{a}})},\ \Eprint
  {http://arxiv.org/abs/hep-lat/0501019} {arXiv:hep-lat/0501019 [hep-lat]}
  \BibitemShut {NoStop}%
\bibitem [{\citenamefont {Furui}\ and\ \citenamefont
  {Nakajima}(2005)}]{Furui:2005mp}%
  \BibitemOpen
  \bibfield  {author} {\bibinfo {author} {\bibfnamefont {S.}~\bibnamefont
  {Furui}}\ and\ \bibinfo {author} {\bibfnamefont {H.}~\bibnamefont
  {Nakajima}},\ }\href@noop {} {\  (\bibinfo {year} {2005})},\ \Eprint
  {http://arxiv.org/abs/hep-lat/0511045} {arXiv:hep-lat/0511045 [hep-lat]}
  \BibitemShut {NoStop}%
\bibitem [{\citenamefont {Neuberger}(1998{\natexlab{a}})}]{Neuberger:1997fp}%
  \BibitemOpen
  \bibfield  {author} {\bibinfo {author} {\bibfnamefont {H.}~\bibnamefont
  {Neuberger}},\ }\href {\doibase 10.1016/S0370-2693(97)01368-3} {\bibfield
  {journal} {\bibinfo  {journal} {Phys. Lett.}\ }\textbf {\bibinfo {volume}
  {417}},\ \bibinfo {pages} {B141} (\bibinfo {year} {1998}{\natexlab{a}})},\
  \Eprint {http://arxiv.org/abs/hep-lat/9707022} {arXiv:hep-lat/9707022}
  \BibitemShut {NoStop}%
\bibitem [{\citenamefont {Neuberger}(1998{\natexlab{b}})}]{Neuberger:1998wv}%
  \BibitemOpen
  \bibfield  {author} {\bibinfo {author} {\bibfnamefont {H.}~\bibnamefont
  {Neuberger}},\ }\href {\doibase 10.1016/S0370-2693(98)00355-4} {\bibfield
  {journal} {\bibinfo  {journal} {Phys. Lett.}\ }\textbf {\bibinfo {volume}
  {427}},\ \bibinfo {pages} {B353} (\bibinfo {year} {1998}{\natexlab{b}})},\
  \Eprint {http://arxiv.org/abs/hep-lat/9801031} {arXiv:hep-lat/9801031}
  \BibitemShut {NoStop}%
\bibitem [{\citenamefont {Bonnet}\ \emph {et~al.}(2002)\citenamefont {Bonnet},
  \citenamefont {Bowman}, \citenamefont {Leinweber}, \citenamefont {Williams},\
  and\ \citenamefont {Zhang}}]{Bonnet:2002ih}%
  \BibitemOpen
  \bibfield  {author} {\bibinfo {author} {\bibfnamefont {F.~D.}\ \bibnamefont
  {Bonnet}}, \bibinfo {author} {\bibfnamefont {P.~O.}\ \bibnamefont {Bowman}},
  \bibinfo {author} {\bibfnamefont {D.~B.}\ \bibnamefont {Leinweber}}, \bibinfo
  {author} {\bibfnamefont {A.~G.}\ \bibnamefont {Williams}}, \ and\ \bibinfo
  {author} {\bibfnamefont {J.-b.}\ \bibnamefont {Zhang}} (\bibinfo
  {collaboration} {CSSM Lattice collaboration}),\ }\href {\doibase
  10.1103/PhysRevD.65.114503} {\bibfield  {journal} {\bibinfo  {journal}
  {Phys.Rev.}\ }\textbf {\bibinfo {volume} {D65}},\ \bibinfo {pages} {114503}
  (\bibinfo {year} {2002})},\ \Eprint {http://arxiv.org/abs/hep-lat/0202003}
  {arXiv:hep-lat/0202003 [hep-lat]} \BibitemShut {NoStop}%
\bibitem [{\citenamefont {Zhang}\ \emph {et~al.}(2004)\citenamefont {Zhang},
  \citenamefont {Bowman}, \citenamefont {Leinweber}, \citenamefont {Williams},\
  and\ \citenamefont {Bonnet}}]{Zhang:2003faa}%
  \BibitemOpen
  \bibfield  {author} {\bibinfo {author} {\bibfnamefont {J.}~\bibnamefont
  {Zhang}}, \bibinfo {author} {\bibfnamefont {P.~O.}\ \bibnamefont {Bowman}},
  \bibinfo {author} {\bibfnamefont {D.~B.}\ \bibnamefont {Leinweber}}, \bibinfo
  {author} {\bibfnamefont {A.~G.}\ \bibnamefont {Williams}}, \ and\ \bibinfo
  {author} {\bibfnamefont {F.~D.}\ \bibnamefont {Bonnet}} (\bibinfo
  {collaboration} {CSSM Lattice collaboration}),\ }\href {\doibase
  10.1103/PhysRevD.70.034505} {\bibfield  {journal} {\bibinfo  {journal}
  {Phys.Rev.}\ }\textbf {\bibinfo {volume} {D70}},\ \bibinfo {pages} {034505}
  (\bibinfo {year} {2004})},\ \Eprint {http://arxiv.org/abs/hep-lat/0301018}
  {arXiv:hep-lat/0301018 [hep-lat]} \BibitemShut {NoStop}%
\bibitem [{\citenamefont {Zhang}\ \emph {et~al.}(2005)\citenamefont {Zhang},
  \citenamefont {Bowman}, \citenamefont {Coad}, \citenamefont {Heller},
  \citenamefont {Leinweber} \emph {et~al.}}]{Zhang:2004gv}%
  \BibitemOpen
  \bibfield  {author} {\bibinfo {author} {\bibfnamefont {J.}~\bibnamefont
  {Zhang}}, \bibinfo {author} {\bibfnamefont {P.~O.}\ \bibnamefont {Bowman}},
  \bibinfo {author} {\bibfnamefont {R.~J.}\ \bibnamefont {Coad}}, \bibinfo
  {author} {\bibfnamefont {U.~M.}\ \bibnamefont {Heller}}, \bibinfo {author}
  {\bibfnamefont {D.~B.}\ \bibnamefont {Leinweber}},  \emph {et~al.},\ }\href
  {\doibase 10.1103/PhysRevD.71.014501} {\bibfield  {journal} {\bibinfo
  {journal} {Phys.Rev.}\ }\textbf {\bibinfo {volume} {D71}},\ \bibinfo {pages}
  {014501} (\bibinfo {year} {2005})},\ \Eprint
  {http://arxiv.org/abs/hep-lat/0410045} {arXiv:hep-lat/0410045 [hep-lat]}
  \BibitemShut {NoStop}%
\bibitem [{\citenamefont {Kamleh}\ \emph {et~al.}(2005)\citenamefont {Kamleh},
  \citenamefont {Bowman}, \citenamefont {Leinweber}, \citenamefont {Williams},\
  and\ \citenamefont {Zhang}}]{Kamleh:2004aw}%
  \BibitemOpen
  \bibfield  {author} {\bibinfo {author} {\bibfnamefont {W.}~\bibnamefont
  {Kamleh}}, \bibinfo {author} {\bibfnamefont {P.~O.}\ \bibnamefont {Bowman}},
  \bibinfo {author} {\bibfnamefont {D.~B.}\ \bibnamefont {Leinweber}}, \bibinfo
  {author} {\bibfnamefont {A.~G.}\ \bibnamefont {Williams}}, \ and\ \bibinfo
  {author} {\bibfnamefont {J.}~\bibnamefont {Zhang}},\ }\href {\doibase
  10.1103/PhysRevD.71.094507} {\bibfield  {journal} {\bibinfo  {journal}
  {Phys.Rev.}\ }\textbf {\bibinfo {volume} {D71}},\ \bibinfo {pages} {094507}
  (\bibinfo {year} {2005})},\ \Eprint {http://arxiv.org/abs/hep-lat/0412022}
  {arXiv:hep-lat/0412022 [hep-lat]} \BibitemShut {NoStop}%
\bibitem [{\citenamefont {Kamleh}\ \emph {et~al.}(2007)\citenamefont {Kamleh},
  \citenamefont {Bowman}, \citenamefont {Leinweber}, \citenamefont {Williams},\
  and\ \citenamefont {Zhang}}]{Kamleh:2007ud}%
  \BibitemOpen
  \bibfield  {author} {\bibinfo {author} {\bibfnamefont {W.}~\bibnamefont
  {Kamleh}}, \bibinfo {author} {\bibfnamefont {P.~O.}\ \bibnamefont {Bowman}},
  \bibinfo {author} {\bibfnamefont {D.~B.}\ \bibnamefont {Leinweber}}, \bibinfo
  {author} {\bibfnamefont {A.~G.}\ \bibnamefont {Williams}}, \ and\ \bibinfo
  {author} {\bibfnamefont {J.}~\bibnamefont {Zhang}},\ }\href {\doibase
  10.1103/PhysRevD.76.094501} {\bibfield  {journal} {\bibinfo  {journal}
  {Phys.Rev.}\ }\textbf {\bibinfo {volume} {D76}},\ \bibinfo {pages} {094501}
  (\bibinfo {year} {2007})},\ \Eprint {http://arxiv.org/abs/0705.4129}
  {arXiv:0705.4129 [hep-lat]} \BibitemShut {NoStop}%
\bibitem [{\citenamefont {Bowman}\ \emph
  {et~al.}(2005{\natexlab{b}})\citenamefont {Bowman}, \citenamefont {Heller},
  \citenamefont {Leinweber}, \citenamefont {Williams},\ and\ \citenamefont
  {Zhang}}]{Bowman:2005zi}%
  \BibitemOpen
  \bibfield  {author} {\bibinfo {author} {\bibfnamefont {P.~O.}\ \bibnamefont
  {Bowman}}, \bibinfo {author} {\bibfnamefont {U.~M.}\ \bibnamefont {Heller}},
  \bibinfo {author} {\bibfnamefont {D.~B.}\ \bibnamefont {Leinweber}}, \bibinfo
  {author} {\bibfnamefont {A.~G.}\ \bibnamefont {Williams}}, \ and\ \bibinfo
  {author} {\bibfnamefont {J.~B.}\ \bibnamefont {Zhang}},\ }\href {\doibase
  10.1007/11356462_2} {\bibfield  {journal} {\bibinfo  {journal} {Lect. Notes
  Phys.}\ }\textbf {\bibinfo {volume} {663}},\ \bibinfo {pages} {17} (\bibinfo
  {year} {2005}{\natexlab{b}})}\BibitemShut {NoStop}%
\bibitem [{\citenamefont {Bowman}\ \emph {et~al.}(2004)\citenamefont {Bowman},
  \citenamefont {Heller}, \citenamefont {Leinweber}, \citenamefont {Williams},\
  and\ \citenamefont {Zhang}}]{Bowman:2004xi}%
  \BibitemOpen
  \bibfield  {author} {\bibinfo {author} {\bibfnamefont {P.~O.}\ \bibnamefont
  {Bowman}}, \bibinfo {author} {\bibfnamefont {U.~M.}\ \bibnamefont {Heller}},
  \bibinfo {author} {\bibfnamefont {D.~B.}\ \bibnamefont {Leinweber}}, \bibinfo
  {author} {\bibfnamefont {A.~G.}\ \bibnamefont {Williams}}, \ and\ \bibinfo
  {author} {\bibfnamefont {J.-b.}\ \bibnamefont {Zhang}},\ }\href {\doibase
  10.1016/S0920-5632(03)02454-X} {\bibfield  {journal} {\bibinfo  {journal}
  {Nucl. Phys. Proc. Suppl.}\ }\textbf {\bibinfo {volume} {128}},\ \bibinfo
  {pages} {23} (\bibinfo {year} {2004})},\ \Eprint
  {http://arxiv.org/abs/hep-lat/0403002} {arXiv:hep-lat/0403002} \BibitemShut
  {NoStop}%
\bibitem [{\citenamefont {Gattringer}(2001)}]{Gattringer:2000js}%
  \BibitemOpen
  \bibfield  {author} {\bibinfo {author} {\bibfnamefont {C.}~\bibnamefont
  {Gattringer}},\ }\href {\doibase 10.1103/PhysRevD.63.114501} {\bibfield
  {journal} {\bibinfo  {journal} {Phys. Rev.}\ }\textbf {\bibinfo {volume}
  {D63}},\ \bibinfo {pages} {114501} (\bibinfo {year} {2001})},\ \Eprint
  {http://arxiv.org/abs/hep-lat/0003005} {arXiv:hep-lat/0003005} \BibitemShut
  {NoStop}%
\bibitem [{\citenamefont {Gattringer}\ \emph {et~al.}(2001)\citenamefont
  {Gattringer}, \citenamefont {Hip},\ and\ \citenamefont
  {Lang}}]{Gattringer:2000qu}%
  \BibitemOpen
  \bibfield  {author} {\bibinfo {author} {\bibfnamefont {C.}~\bibnamefont
  {Gattringer}}, \bibinfo {author} {\bibfnamefont {I.}~\bibnamefont {Hip}}, \
  and\ \bibinfo {author} {\bibfnamefont {C.~B.}\ \bibnamefont {Lang}},\ }\href
  {\doibase 10.1016/S0550-3213(00)00717-3} {\bibfield  {journal} {\bibinfo
  {journal} {Nucl. Phys.}\ }\textbf {\bibinfo {volume} {597}},\ \bibinfo
  {pages} {B451} (\bibinfo {year} {2001})},\ \Eprint
  {http://arxiv.org/abs/hep-lat/0007042} {arXiv:hep-lat/0007042} \BibitemShut
  {NoStop}%
\bibitem [{\citenamefont {Gattringer}\ \emph {et~al.}(2009)\citenamefont
  {Gattringer}, \citenamefont {Hagen}, \citenamefont {Lang}, \citenamefont
  {Limmer}, \citenamefont {Mohler},\ and\ \citenamefont
  {Sch{\"a}fer}}]{Gattringer:2008vj}%
  \BibitemOpen
  \bibfield  {author} {\bibinfo {author} {\bibfnamefont {C.}~\bibnamefont
  {Gattringer}}, \bibinfo {author} {\bibfnamefont {C.}~\bibnamefont {Hagen}},
  \bibinfo {author} {\bibfnamefont {C.~B.}\ \bibnamefont {Lang}}, \bibinfo
  {author} {\bibfnamefont {M.}~\bibnamefont {Limmer}}, \bibinfo {author}
  {\bibfnamefont {D.}~\bibnamefont {Mohler}}, \ and\ \bibinfo {author}
  {\bibfnamefont {A.}~\bibnamefont {Sch{\"a}fer}},\ }\href {\doibase
  10.1103/PhysRevD.79.054501} {\bibfield  {journal} {\bibinfo  {journal} {Phys.
  Rev.}\ }\textbf {\bibinfo {volume} {D79}},\ \bibinfo {pages} {054501}
  (\bibinfo {year} {2009})},\ \Eprint {http://arxiv.org/abs/0812.1681}
  {arXiv:0812.1681 [hep-lat]} \BibitemShut {NoStop}%
\bibitem [{\citenamefont {Engel}\ \emph {et~al.}(2010)\citenamefont {Engel},
  \citenamefont {Lang}, \citenamefont {Limmer}, \citenamefont {Mohler},\ and\
  \citenamefont {Sch{\"a}fer}}]{Engel:2010my}%
  \BibitemOpen
  \bibfield  {author} {\bibinfo {author} {\bibfnamefont {G.~P.}\ \bibnamefont
  {Engel}}, \bibinfo {author} {\bibfnamefont {C.~B.}\ \bibnamefont {Lang}},
  \bibinfo {author} {\bibfnamefont {M.}~\bibnamefont {Limmer}}, \bibinfo
  {author} {\bibfnamefont {D.}~\bibnamefont {Mohler}}, \ and\ \bibinfo {author}
  {\bibfnamefont {A.}~\bibnamefont {Sch{\"a}fer}} (\bibinfo {collaboration}
  {BGR [Bern-Graz-Regensburg]}),\ }\href@noop {} {\bibfield  {journal}
  {\bibinfo  {journal} {Phys. Rev.}\ }\textbf {\bibinfo {volume} {D82}},\
  \bibinfo {pages} {034505} (\bibinfo {year} {2010})},\ \Eprint
  {http://arxiv.org/abs/1005.1748} {arXiv:1005.1748 [hep-lat]} \BibitemShut
  {NoStop}%
\bibitem [{\citenamefont {Banks}\ and\ \citenamefont
  {Casher}(1980)}]{Banks:1979yr}%
  \BibitemOpen
  \bibfield  {author} {\bibinfo {author} {\bibfnamefont {T.}~\bibnamefont
  {Banks}}\ and\ \bibinfo {author} {\bibfnamefont {A.}~\bibnamefont {Casher}},\
  }\href {\doibase 10.1016/0550-3213(80)90255-2} {\bibfield  {journal}
  {\bibinfo  {journal} {Nucl. Phys.}\ }\textbf {\bibinfo {volume} {169}},\
  \bibinfo {pages} {B103} (\bibinfo {year} {1980})}\BibitemShut {NoStop}%
\bibitem [{\citenamefont {{Lang, C.B. and Schr{\"o}ck,
  Mario}}(2011)}]{Lang:2011vw}%
  \BibitemOpen
  \bibfield  {author} {\bibinfo {author} {\bibnamefont {{Lang, C.B. and
  Schr{\"o}ck, Mario}}},\ }\href {\doibase 10.1103/PhysRevD.84.087704}
  {\bibfield  {journal} {\bibinfo  {journal} {Phys.Rev.}\ }\textbf {\bibinfo
  {volume} {D84}},\ \bibinfo {pages} {087704} (\bibinfo {year} {2011})},\
  \Eprint {http://arxiv.org/abs/1107.5195} {arXiv:1107.5195 [hep-lat]}
  \BibitemShut {NoStop}%
\bibitem [{\citenamefont {Glozman}(2007)}]{Glozman:2007ek}%
  \BibitemOpen
  \bibfield  {author} {\bibinfo {author} {\bibfnamefont {L.}~\bibnamefont
  {Glozman}},\ }\href {\doibase 10.1016/j.physrep.2007.04.001} {\bibfield
  {journal} {\bibinfo  {journal} {Phys.Rept.}\ }\textbf {\bibinfo {volume}
  {444}},\ \bibinfo {pages} {1} (\bibinfo {year} {2007})},\ \Eprint
  {http://arxiv.org/abs/hep-ph/0701081} {arXiv:hep-ph/0701081 [hep-ph]}
  \BibitemShut {NoStop}%
\bibitem [{\citenamefont {Suganuma}\ \emph {et~al.}(2011)\citenamefont
  {Suganuma}, \citenamefont {Gongyo}, \citenamefont {Iritani},\ and\
  \citenamefont {Yamamoto}}]{Suganuma:2011kn}%
  \BibitemOpen
  \bibfield  {author} {\bibinfo {author} {\bibfnamefont {H.}~\bibnamefont
  {Suganuma}}, \bibinfo {author} {\bibfnamefont {S.}~\bibnamefont {Gongyo}},
  \bibinfo {author} {\bibfnamefont {T.}~\bibnamefont {Iritani}}, \ and\
  \bibinfo {author} {\bibfnamefont {A.}~\bibnamefont {Yamamoto}},\ }\href@noop
  {} {\  (\bibinfo {year} {2011})},\ \Eprint {http://arxiv.org/abs/1112.1962}
  {arXiv:1112.1962 [hep-lat]} \BibitemShut {NoStop}%
\bibitem [{\citenamefont {Giusti}\ \emph {et~al.}(2001)\citenamefont {Giusti},
  \citenamefont {Paciello}, \citenamefont {Parrinello}, \citenamefont
  {Petrarca},\ and\ \citenamefont {Taglienti}}]{Giusti:2001xf}%
  \BibitemOpen
  \bibfield  {author} {\bibinfo {author} {\bibfnamefont {L.}~\bibnamefont
  {Giusti}}, \bibinfo {author} {\bibfnamefont {M.}~\bibnamefont {Paciello}},
  \bibinfo {author} {\bibfnamefont {C.}~\bibnamefont {Parrinello}}, \bibinfo
  {author} {\bibfnamefont {S.}~\bibnamefont {Petrarca}}, \ and\ \bibinfo
  {author} {\bibfnamefont {B.}~\bibnamefont {Taglienti}},\ }\href {\doibase
  10.1142/S0217751X01004281} {\bibfield  {journal} {\bibinfo  {journal}
  {Int.J.Mod.Phys.}\ }\textbf {\bibinfo {volume} {A16}},\ \bibinfo {pages}
  {3487} (\bibinfo {year} {2001})},\ \Eprint
  {http://arxiv.org/abs/hep-lat/0104012} {arXiv:hep-lat/0104012 [hep-lat]}
  \BibitemShut {NoStop}%
\bibitem [{\citenamefont {Gattringer}\ \emph
  {et~al.}(2004{\natexlab{a}})\citenamefont {Gattringer}, \citenamefont
  {G{\"o}ckeler}, \citenamefont {Hasenfratz}, \citenamefont {Hauswirth},
  \citenamefont {Holland}, \citenamefont {J{\"o}rg}, \citenamefont {Juge},
  \citenamefont {Lang}, \citenamefont {Niedermayer}, \citenamefont {Rakow},
  \citenamefont {Schaefer},\ and\ \citenamefont {Schafer}}]{Gattringer:2003qx}%
  \BibitemOpen
  \bibfield  {author} {\bibinfo {author} {\bibfnamefont {C.}~\bibnamefont
  {Gattringer}}, \bibinfo {author} {\bibfnamefont {M.}~\bibnamefont
  {G{\"o}ckeler}}, \bibinfo {author} {\bibfnamefont {P.}~\bibnamefont
  {Hasenfratz}}, \bibinfo {author} {\bibfnamefont {S.}~\bibnamefont
  {Hauswirth}}, \bibinfo {author} {\bibfnamefont {K.}~\bibnamefont {Holland}},
  \bibinfo {author} {\bibfnamefont {T.}~\bibnamefont {J{\"o}rg}}, \bibinfo
  {author} {\bibfnamefont {K.~J.}\ \bibnamefont {Juge}}, \bibinfo {author}
  {\bibfnamefont {C.~B.}\ \bibnamefont {Lang}}, \bibinfo {author}
  {\bibfnamefont {F.}~\bibnamefont {Niedermayer}}, \bibinfo {author}
  {\bibfnamefont {P.~E.~L.}\ \bibnamefont {Rakow}}, \bibinfo {author}
  {\bibfnamefont {S.}~\bibnamefont {Schaefer}}, \ and\ \bibinfo {author}
  {\bibfnamefont {A.}~\bibnamefont {Schafer}},\ }\href {\doibase
  10.1016/j.nuclphysb.2003.10.044} {\bibfield  {journal} {\bibinfo  {journal}
  {Nucl. Phys.}\ }\textbf {\bibinfo {volume} {B677}},\ \bibinfo {pages} {3}
  (\bibinfo {year} {2004}{\natexlab{a}})},\ \Eprint
  {http://arxiv.org/abs/hep-lat/0307013} {arXiv:hep-lat/0307013} \BibitemShut
  {NoStop}%
\bibitem [{\citenamefont {Lang}\ \emph {et~al.}(2006)\citenamefont {Lang},
  \citenamefont {Majumdar},\ and\ \citenamefont {Ortner}}]{Lang:2005jz}%
  \BibitemOpen
  \bibfield  {author} {\bibinfo {author} {\bibfnamefont {C.~B.}\ \bibnamefont
  {Lang}}, \bibinfo {author} {\bibfnamefont {P.}~\bibnamefont {Majumdar}}, \
  and\ \bibinfo {author} {\bibfnamefont {W.}~\bibnamefont {Ortner}},\ }\href
  {\doibase 10.1103/PhysRevD.73.034507} {\bibfield  {journal} {\bibinfo
  {journal} {Phys. Rev.}\ }\textbf {\bibinfo {volume} {D73}},\ \bibinfo {pages}
  {034507} (\bibinfo {year} {2006})},\ \Eprint
  {http://arxiv.org/abs/hep-lat/0512014} {arXiv:hep-lat/0512014} \BibitemShut
  {NoStop}%
\bibitem [{\citenamefont {{Engel, Georg P. and Lang, C.B. and Limmer, Markus
  and Mohler, Daniel and Schaefer, Andreas}}(2011)}]{Engel:2011zr}%
  \BibitemOpen
  \bibfield  {author} {\bibinfo {author} {\bibnamefont {{Engel, Georg P. and
  Lang, C.B. and Limmer, Markus and Mohler, Daniel and Schaefer, Andreas}}},\
  }\href@noop {} {\  (\bibinfo {year} {2011})},\ \bibinfo {note} {* Temporary
  entry *},\ \Eprint {http://arxiv.org/abs/1112.1601} {arXiv:1112.1601
  [hep-lat]} \BibitemShut {NoStop}%
\bibitem [{\citenamefont {Gattringer}\ \emph
  {et~al.}(2004{\natexlab{b}})\citenamefont {Gattringer}, \citenamefont
  {Gockeler}, \citenamefont {Huber},\ and\ \citenamefont
  {Lang}}]{Gattringer:2004iv}%
  \BibitemOpen
  \bibfield  {author} {\bibinfo {author} {\bibfnamefont {C.}~\bibnamefont
  {Gattringer}}, \bibinfo {author} {\bibfnamefont {M.}~\bibnamefont
  {Gockeler}}, \bibinfo {author} {\bibfnamefont {P.}~\bibnamefont {Huber}}, \
  and\ \bibinfo {author} {\bibfnamefont {C.}~\bibnamefont {Lang}},\ }\href
  {\doibase 10.1016/j.nuclphysb.2004.06.013} {\bibfield  {journal} {\bibinfo
  {journal} {Nucl.Phys.}\ }\textbf {\bibinfo {volume} {B694}},\ \bibinfo
  {pages} {170} (\bibinfo {year} {2004}{\natexlab{b}})},\ \Eprint
  {http://arxiv.org/abs/hep-lat/0404006} {arXiv:hep-lat/0404006 [hep-lat]}
  \BibitemShut {NoStop}%
\bibitem [{\citenamefont {Huber}(2010)}]{Huber:2010zza}%
  \BibitemOpen
  \bibfield  {author} {\bibinfo {author} {\bibfnamefont {P.}~\bibnamefont
  {Huber}},\ }\href {\doibase 10.1007/JHEP11(2010)107} {\bibfield  {journal}
  {\bibinfo  {journal} {JHEP}\ }\textbf {\bibinfo {volume} {1011}},\ \bibinfo
  {pages} {107} (\bibinfo {year} {2010})},\ \Eprint
  {http://arxiv.org/abs/1003.3496} {arXiv:1003.3496 [hep-lat]} \BibitemShut
  {NoStop}%
\bibitem [{\citenamefont {Sheikholeslami}\ and\ \citenamefont
  {Wohlert}(1985{\natexlab{a}})}]{ShWo85}%
  \BibitemOpen
  \bibfield  {author} {\bibinfo {author} {\bibfnamefont {B.}~\bibnamefont
  {Sheikholeslami}}\ and\ \bibinfo {author} {\bibfnamefont {R.}~\bibnamefont
  {Wohlert}},\ }\href@noop {} {\bibfield  {journal} {\bibinfo  {journal} {Nucl.
  Phys.}\ }\textbf {\bibinfo {volume} {B259}},\ \bibinfo {pages} {572}
  (\bibinfo {year} {1985}{\natexlab{a}})}\BibitemShut {NoStop}%
\bibitem [{\citenamefont {Skullerud}\ and\ \citenamefont
  {Williams}(2000)}]{Skullerud:1999gv}%
  \BibitemOpen
  \bibfield  {author} {\bibinfo {author} {\bibfnamefont {J.~I.}\ \bibnamefont
  {Skullerud}}\ and\ \bibinfo {author} {\bibfnamefont {A.~G.}\ \bibnamefont
  {Williams}},\ }\href@noop {} {\bibfield  {journal} {\bibinfo  {journal}
  {Nucl.Phys.Proc.Suppl.}\ }\textbf {\bibinfo {volume} {83}},\ \bibinfo {pages}
  {209} (\bibinfo {year} {2000})},\ \Eprint
  {http://arxiv.org/abs/hep-lat/9909142} {arXiv:hep-lat/9909142 [hep-lat]}
  \BibitemShut {NoStop}%
\bibitem [{\citenamefont {Alkofer}\ and\ \citenamefont {von
  Smekal}(2001)}]{Alkofer:2000wg}%
  \BibitemOpen
  \bibfield  {author} {\bibinfo {author} {\bibfnamefont {R.}~\bibnamefont
  {Alkofer}}\ and\ \bibinfo {author} {\bibfnamefont {L.}~\bibnamefont {von
  Smekal}},\ }\href {\doibase 10.1016/S0370-1573(01)00010-2} {\bibfield
  {journal} {\bibinfo  {journal} {Phys.Rept.}\ }\textbf {\bibinfo {volume}
  {353}},\ \bibinfo {pages} {281} (\bibinfo {year} {2001})},\ \Eprint
  {http://arxiv.org/abs/hep-ph/0007355} {arXiv:hep-ph/0007355 [hep-ph]}
  \BibitemShut {NoStop}%
\bibitem [{\citenamefont {Symanzik}(1983)}]{Symanzik:1983dc}%
  \BibitemOpen
  \bibfield  {author} {\bibinfo {author} {\bibfnamefont {K.}~\bibnamefont
  {Symanzik}},\ }\href {\doibase 10.1016/0550-3213(83)90468-6} {\bibfield
  {journal} {\bibinfo  {journal} {Nucl.Phys.}\ }\textbf {\bibinfo {volume}
  {B226}},\ \bibinfo {pages} {187} (\bibinfo {year} {1983})}\BibitemShut
  {NoStop}%
\bibitem [{\citenamefont {L{\"u}scher}\ \emph {et~al.}(1996)\citenamefont
  {L{\"u}scher}, \citenamefont {Sint}, \citenamefont {Sommer},\ and\
  \citenamefont {Weisz}}]{Luscher:1996sc}%
  \BibitemOpen
  \bibfield  {author} {\bibinfo {author} {\bibfnamefont {M.}~\bibnamefont
  {L{\"u}scher}}, \bibinfo {author} {\bibfnamefont {S.}~\bibnamefont {Sint}},
  \bibinfo {author} {\bibfnamefont {R.}~\bibnamefont {Sommer}}, \ and\ \bibinfo
  {author} {\bibfnamefont {P.}~\bibnamefont {Weisz}},\ }\href {\doibase
  10.1016/0550-3213(96)00378-1} {\bibfield  {journal} {\bibinfo  {journal}
  {Nucl.Phys.}\ }\textbf {\bibinfo {volume} {B478}},\ \bibinfo {pages} {365}
  (\bibinfo {year} {1996})},\ \Eprint {http://arxiv.org/abs/hep-lat/9605038}
  {arXiv:hep-lat/9605038 [hep-lat]} \BibitemShut {NoStop}%
\bibitem [{\citenamefont {Sheikholeslami}\ and\ \citenamefont
  {Wohlert}(1985{\natexlab{b}})}]{Sheikholeslami:1985ij}%
  \BibitemOpen
  \bibfield  {author} {\bibinfo {author} {\bibfnamefont {B.}~\bibnamefont
  {Sheikholeslami}}\ and\ \bibinfo {author} {\bibfnamefont {R.}~\bibnamefont
  {Wohlert}},\ }\href {\doibase 10.1016/0550-3213(85)90002-1} {\bibfield
  {journal} {\bibinfo  {journal} {Nucl.Phys.}\ }\textbf {\bibinfo {volume}
  {B259}},\ \bibinfo {pages} {572} (\bibinfo {year}
  {1985}{\natexlab{b}})}\BibitemShut {NoStop}%
\bibitem [{\citenamefont {Heatlie}\ \emph {et~al.}(1991)\citenamefont
  {Heatlie}, \citenamefont {Martinelli}, \citenamefont {Pittori}, \citenamefont
  {Rossi},\ and\ \citenamefont {Sachrajda}}]{Heatlie:1990kg}%
  \BibitemOpen
  \bibfield  {author} {\bibinfo {author} {\bibfnamefont {G.}~\bibnamefont
  {Heatlie}}, \bibinfo {author} {\bibfnamefont {G.}~\bibnamefont {Martinelli}},
  \bibinfo {author} {\bibfnamefont {C.}~\bibnamefont {Pittori}}, \bibinfo
  {author} {\bibfnamefont {G.}~\bibnamefont {Rossi}}, \ and\ \bibinfo {author}
  {\bibfnamefont {C.~T.}\ \bibnamefont {Sachrajda}},\ }\href {\doibase
  10.1016/0550-3213(91)90137-M} {\bibfield  {journal} {\bibinfo  {journal}
  {Nucl.Phys.}\ }\textbf {\bibinfo {volume} {B352}},\ \bibinfo {pages} {266}
  (\bibinfo {year} {1991})}\BibitemShut {NoStop}%
\bibitem [{\citenamefont {Dawson}\ \emph {et~al.}(1998)\citenamefont {Dawson},
  \citenamefont {Martinelli}, \citenamefont {Rossi}, \citenamefont {Sachrajda},
  \citenamefont {Sharpe} \emph {et~al.}}]{Dawson:1997gp}%
  \BibitemOpen
  \bibfield  {author} {\bibinfo {author} {\bibfnamefont {C.}~\bibnamefont
  {Dawson}}, \bibinfo {author} {\bibfnamefont {G.}~\bibnamefont {Martinelli}},
  \bibinfo {author} {\bibfnamefont {G.}~\bibnamefont {Rossi}}, \bibinfo
  {author} {\bibfnamefont {C.~T.}\ \bibnamefont {Sachrajda}}, \bibinfo {author}
  {\bibfnamefont {S.~R.}\ \bibnamefont {Sharpe}},  \emph {et~al.},\ }\href
  {\doibase 10.1016/S0920-5632(97)00927-4} {\bibfield  {journal} {\bibinfo
  {journal} {Nucl.Phys.Proc.Suppl.}\ }\textbf {\bibinfo {volume} {63}},\
  \bibinfo {pages} {877} (\bibinfo {year} {1998})},\ \Eprint
  {http://arxiv.org/abs/hep-lat/9710027} {arXiv:hep-lat/9710027 [hep-lat]}
  \BibitemShut {NoStop}%
\bibitem [{\citenamefont {Egri}\ \emph {et~al.}(2007)\citenamefont {Egri},
  \citenamefont {Fodor}, \citenamefont {Hoelbling}, \citenamefont {Katz},
  \citenamefont {Nogradi} \emph {et~al.}}]{Egri:2006zm}%
  \BibitemOpen
  \bibfield  {author} {\bibinfo {author} {\bibfnamefont {G.~I.}\ \bibnamefont
  {Egri}}, \bibinfo {author} {\bibfnamefont {Z.}~\bibnamefont {Fodor}},
  \bibinfo {author} {\bibfnamefont {C.}~\bibnamefont {Hoelbling}}, \bibinfo
  {author} {\bibfnamefont {S.~D.}\ \bibnamefont {Katz}}, \bibinfo {author}
  {\bibfnamefont {D.}~\bibnamefont {Nogradi}},  \emph {et~al.},\ }\href
  {\doibase 10.1016/j.cpc.2007.06.005} {\bibfield  {journal} {\bibinfo
  {journal} {Comput.Phys.Commun.}\ }\textbf {\bibinfo {volume} {177}},\
  \bibinfo {pages} {631} (\bibinfo {year} {2007})},\ \Eprint
  {http://arxiv.org/abs/hep-lat/0611022} {arXiv:hep-lat/0611022 [hep-lat]}
  \BibitemShut {NoStop}%
\bibitem [{\citenamefont {Babich}\ \emph {et~al.}(2010)\citenamefont {Babich},
  \citenamefont {Clark},\ and\ \citenamefont {Joo}}]{Babich:2010mu}%
  \BibitemOpen
  \bibfield  {author} {\bibinfo {author} {\bibfnamefont {R.}~\bibnamefont
  {Babich}}, \bibinfo {author} {\bibfnamefont {M.~A.}\ \bibnamefont {Clark}}, \
  and\ \bibinfo {author} {\bibfnamefont {B.}~\bibnamefont {Joo}},\ }\href@noop
  {} {\  (\bibinfo {year} {2010})},\ \Eprint {http://arxiv.org/abs/1011.0024}
  {arXiv:1011.0024 [hep-lat]} \BibitemShut {NoStop}%
\bibitem [{\citenamefont {Clark}\ \emph {et~al.}(2010)\citenamefont {Clark},
  \citenamefont {Babich}, \citenamefont {Barros}, \citenamefont {Brower},\ and\
  \citenamefont {Rebbi}}]{Clark:2009wm}%
  \BibitemOpen
  \bibfield  {author} {\bibinfo {author} {\bibfnamefont {M.~A.}\ \bibnamefont
  {Clark}}, \bibinfo {author} {\bibfnamefont {R.}~\bibnamefont {Babich}},
  \bibinfo {author} {\bibfnamefont {K.}~\bibnamefont {Barros}}, \bibinfo
  {author} {\bibfnamefont {R.~C.}\ \bibnamefont {Brower}}, \ and\ \bibinfo
  {author} {\bibfnamefont {C.}~\bibnamefont {Rebbi}},\ }\href {\doibase
  10.1016/j.cpc.2010.05.002} {\bibfield  {journal} {\bibinfo  {journal}
  {Comput. Phys. Commun.}\ }\textbf {\bibinfo {volume} {181}},\ \bibinfo
  {pages} {1517} (\bibinfo {year} {2010})},\ \Eprint
  {http://arxiv.org/abs/0911.3191} {arXiv:0911.3191 [hep-lat]} \BibitemShut
  {NoStop}%
\bibitem [{\citenamefont {Winter}(2011)}]{Winter:2011dh}%
  \BibitemOpen
  \bibfield  {author} {\bibinfo {author} {\bibfnamefont {F.}~\bibnamefont
  {Winter}},\ }\href@noop {} {\  (\bibinfo {year} {2011})},\ \Eprint
  {http://arxiv.org/abs/1111.5596} {arXiv:1111.5596 [hep-lat]} \BibitemShut
  {NoStop}%
\bibitem [{\citenamefont {Lujan}\ \emph {et~al.}(2011)\citenamefont {Lujan},
  \citenamefont {Alexandru},\ and\ \citenamefont {Lee}}]{Lujan:2011ue}%
  \BibitemOpen
  \bibfield  {author} {\bibinfo {author} {\bibfnamefont {M.}~\bibnamefont
  {Lujan}}, \bibinfo {author} {\bibfnamefont {A.}~\bibnamefont {Alexandru}}, \
  and\ \bibinfo {author} {\bibfnamefont {F.}~\bibnamefont {Lee}},\ }\href@noop
  {} {\  (\bibinfo {year} {2011})},\ \Eprint {http://arxiv.org/abs/1111.6288}
  {arXiv:1111.6288 [hep-lat]} \BibitemShut {NoStop}%
\bibitem [{\citenamefont {Cardoso}\ and\ \citenamefont
  {Bicudo}(2011{\natexlab{a}})}]{Cardoso:2011xu}%
  \BibitemOpen
  \bibfield  {author} {\bibinfo {author} {\bibfnamefont {N.}~\bibnamefont
  {Cardoso}}\ and\ \bibinfo {author} {\bibfnamefont {P.}~\bibnamefont
  {Bicudo}},\ }\href@noop {} {\  (\bibinfo {year} {2011}{\natexlab{a}})},\
  \Eprint {http://arxiv.org/abs/1112.4533} {arXiv:1112.4533 [hep-lat]}
  \BibitemShut {NoStop}%
\bibitem [{\citenamefont {Jang}\ \emph {et~al.}(2011)\citenamefont {Jang},
  \citenamefont {Kim},\ and\ \citenamefont {Lee}}]{Jang:2011cn}%
  \BibitemOpen
  \bibfield  {author} {\bibinfo {author} {\bibfnamefont {Y.-C.}\ \bibnamefont
  {Jang}}, \bibinfo {author} {\bibfnamefont {H.-J.}\ \bibnamefont {Kim}}, \
  and\ \bibinfo {author} {\bibfnamefont {W.}~\bibnamefont {Lee}},\ }\href@noop
  {} {\bibfield  {journal} {\bibinfo  {journal} {PoS}\ }\textbf {\bibinfo
  {volume} {LATTICE2011}},\ \bibinfo {pages} {309} (\bibinfo {year} {2011})},\
  \Eprint {http://arxiv.org/abs/1111.0125} {arXiv:1111.0125 [physics.comp-ph]}
  \BibitemShut {NoStop}%
\bibitem [{\citenamefont {Walk}\ \emph {et~al.}(2010)\citenamefont {Walk},
  \citenamefont {Wittig}, \citenamefont {Dranischnikow},\ and\ \citenamefont
  {Schomer}}]{Walk:2010ut}%
  \BibitemOpen
  \bibfield  {author} {\bibinfo {author} {\bibfnamefont {B.}~\bibnamefont
  {Walk}}, \bibinfo {author} {\bibfnamefont {H.}~\bibnamefont {Wittig}},
  \bibinfo {author} {\bibfnamefont {E.}~\bibnamefont {Dranischnikow}}, \ and\
  \bibinfo {author} {\bibfnamefont {E.}~\bibnamefont {Schomer}},\ }\href@noop
  {} {\bibfield  {journal} {\bibinfo  {journal} {PoS}\ }\textbf {\bibinfo
  {volume} {LATTICE2010}},\ \bibinfo {pages} {044} (\bibinfo {year} {2010})},\
  \Eprint {http://arxiv.org/abs/1010.5636} {arXiv:1010.5636 [hep-lat]}
  \BibitemShut {NoStop}%
\bibitem [{\citenamefont {Cardoso}\ and\ \citenamefont
  {Bicudo}(2011{\natexlab{b}})}]{Cardoso:2010di}%
  \BibitemOpen
  \bibfield  {author} {\bibinfo {author} {\bibfnamefont {N.}~\bibnamefont
  {Cardoso}}\ and\ \bibinfo {author} {\bibfnamefont {P.}~\bibnamefont
  {Bicudo}},\ }\href {\doibase 10.1016/j.jcp.2011.02.023} {\bibfield  {journal}
  {\bibinfo  {journal} {J.Comput.Phys.}\ }\textbf {\bibinfo {volume} {230}},\
  \bibinfo {pages} {3998} (\bibinfo {year} {2011}{\natexlab{b}})},\ \Eprint
  {http://arxiv.org/abs/1010.4834} {arXiv:1010.4834 [hep-lat]} \BibitemShut
  {NoStop}%
\bibitem [{\citenamefont {Mandula}\ and\ \citenamefont
  {Ogilvie}(1987)}]{Mandula:1987rh}%
  \BibitemOpen
  \bibfield  {author} {\bibinfo {author} {\bibfnamefont {J.~E.}\ \bibnamefont
  {Mandula}}\ and\ \bibinfo {author} {\bibfnamefont {M.}~\bibnamefont
  {Ogilvie}},\ }\href {\doibase 10.1016/0370-2693(87)91541-3} {\bibfield
  {journal} {\bibinfo  {journal} {Phys. Lett.}\ }\textbf {\bibinfo {volume}
  {B185}},\ \bibinfo {pages} {127} (\bibinfo {year} {1987})}\BibitemShut
  {NoStop}%
\bibitem [{\citenamefont {Mandula}\ and\ \citenamefont
  {Ogilvie}(1990)}]{Mandula:1990vs}%
  \BibitemOpen
  \bibfield  {author} {\bibinfo {author} {\bibfnamefont {J.~E.}\ \bibnamefont
  {Mandula}}\ and\ \bibinfo {author} {\bibfnamefont {M.}~\bibnamefont
  {Ogilvie}},\ }\href {\doibase 10.1016/0370-2693(90)90031-Z} {\bibfield
  {journal} {\bibinfo  {journal} {Phys.Lett.}\ }\textbf {\bibinfo {volume}
  {B248}},\ \bibinfo {pages} {156} (\bibinfo {year} {1990})}\BibitemShut
  {NoStop}%
\end{thebibliography}

\end{document}